\documentclass[12pt,preprint]{aastex}

\usepackage{graphicx}
\usepackage{subfigure}
\usepackage{subfloat}
\usepackage{amssymb}
\usepackage{amsmath}
\usepackage{multirow}
\usepackage{dcolumn}
\usepackage{bm}

\shorttitle{Collapsed Cores in Globular Clusters}
\shortauthors{}

\begin{document}

\title{Evolution of Kinetic and Magnetic Energy in Intra Cluster Media}

\author{Kiwan Park}
\affil{Department of Physics, UNIST, Ulsan, 689798, Korea; pkiwan@unist.ac.kr}

\author{Dongho Park}
\affil{APCPT, Pohang, 790784, Korea;  dongho@apctp.org}

%
%
%

\begin{abstract}
Intra Cluster Media (ICMs) located at galaxy clusters is in the state of hot, tenuous, magnetized, and highly ionized X-ray emitting plasmas. This overall collisionless, viscous, and conductive magnetohydrodynamic (MHD) turbulence in ICM is simulated using hyper and physical magnetic diffusivity. The results show that fluctuating random plasma motion amplifies the magnetic field, which cascades toward the diffusivity scale passing through the viscous scale. The kinetic eddies in the subviscous scale are driven and constrained by the magnetic tension which finally gets balanced with the highly damping effect of the kinetic eddies. However, the saturated kinetic energy spectrum is deeper than that of the incompressible or compressible hydrodynamics fluid. To explain this unusual field profile we set up two simultaneous differential equations for the kinetic and magnetic energy spectrum using an Eddy Damped Quasi Normal Markovianized (EDQNM) approximation. The analytic solution tells us that the magnetic energy in addition to the viscous damping effect constrains the plasma motion leading to the power spectra: kinetic energy spectrum $E_V^k\sim k^{-3}$ and corresponding representative magnetic energy spectrum $E_M^k\sim k^{-1/2}$. Also the comparison of simulation results with different resolutions and magnetic diffusivities implies the role of small scale magnetic energy in dynamo.
\end{abstract}

\keywords{galaxies: clusters: intracluster medium, magnetic fields}

\section{Introduction}
ICM located at the center of galaxy cluster is composed of fully ionized hot plasmas ($T\sim10^8$K). The gas includes most of the cluster baryons ($>85\%$) and heavy elements, but overall density `$n$' is very low ($n<10^3\, cm^{-3}$). As a result ICM has very small diffusivity `$\eta$' ($\sim T^{-3/2}$) while viscosity `$\nu$' ($\sim 1/n$) is very large \citep{Schekochihin et al 2005}. Dynamo effect from the turbulent flow motions in ICM exceeds the dissipation of such high viscous plasmas causing the growth of seed magnetic fields, which react back the plasma motion through magnetic tension ($\mathbf{B}\cdot \nabla \mathbf{B}$).\\

\noindent The amplified magnetic field constrains some intrinsic properties in ICM. Non-magnetized thermal Spitzer conductivity `$\kappa_{SP}$' is like \citep{Narayan and Medvedev 2001}:
\begin{eqnarray}
\kappa_{SP} \sim \frac{\lambda_{\mathrm{e}}^2}{t_{coul}}\sim 4\times 10^{32}T_{1}^{5/2}n^{-1}_{-3}\,\mathrm{cm}^2\,s^{-1},
\label{SPConductivity}
\end{eqnarray}
where $T_1=kT/10$ Kev$, \lambda_{\mathrm{e}}$ is the mean free path of an electron and $t_{coul}$ is the time between coulomb collisions. If magnetic field is injected, $\kappa_\bot$, conductivity perpendicular to the magnetic field is reduced to $\sim(\rho_e/\lambda_e)^2\kappa_{SP}$($\ll \kappa_{SP}$) so that the conductivity becomes anisotropic with one third of the thermal Spitzer conductivity: $\kappa=\kappa_\bot+\kappa_\|\rightarrow \kappa_\|$. This anisotropic conductivity brings about temperature distribution that depends on the direction and strength of magnetic field. Also as the conservation of magnetic moment $\mu_B=u^2_\bot/B$ and kinetic energy ($\sim u^2_\|+u^2_\bot$) implies, magnetic field makes pressure tensor $P_{ij}=mn\overline{u_iu_j}$ anisotropic along the field. The ratio of pressure anisotropy `$\triangle$' is implicitly related to the viscosity and stability of  ICM \citep{Schekochihin et al 2010}:
\begin{eqnarray}
\triangle&\equiv&\frac{p_\bot-p_\|}{p}\sim\frac{1}{\nu_{ii}}\frac{1}{B}\frac{dB}{dt}\in \bigg[-\frac{2}{\beta}, \frac{1}{\beta}\bigg],\qquad (\beta=8\pi p/B^2).
\label{pressure_def}
\end{eqnarray}
Here, kinematic viscosity $\nu\sim u_{th}^2/\nu_{ii}$, ion-ion collision frequency $\nu_{ii}=4\pi ne^4\textmd{ln}\,\Lambda\, m_i^{-1/2}T^{3/2}$. If $\triangle$ is smaller than $-2/\beta$ or larger than $1/\beta$, firehose or mirror instability occurs in the range between ion Larmor radius ($\rho_i\sim 10^4-10^6$ km) and the mean free path ($\mathrm{\lambda_{mfp}}$$\sim 10^{15}$ km) \citep{Schekochihin et al 2008}. Since ICM has high `$\beta$', stability range is very narrow, i.e., practically unstable. Instability is thought to redistribute the plasma motions faster than coulomb collision does, but its role is not fully understood yet \citep{Jones 2008}.\\

\noindent By now we may begin to wonder if a typical MHD theory is applicable to the weakly collisional ICM plasma. Since collision between particles transfers momentum to make the system isotropic, most phenomenological or analytic methods assume the collisional MHD system system. However, since the tenuous plasmas cannot expect sufficient collisions, Chew, Goldberger, and Low \citep{Chew et al 1956} developed an adiabatic MHD equation with the consideration of anisotropic pressure tensor (CGL-MHD model). Recently Santos-Liman insisted the validness of the typical MHD theory with these additional constraints \citep{Santos-Liman et al 2011, Santos-Liman et al 2014}. Their model, a kinetic MHD based on CGL-closure with the limit of anisotropy, shows the growth rate of magnetic energy is similar to that of a typical MHD theory in the early time regime, but the saturation is much smaller. In fact, the effect of anisotropy due to the magnetic field in the plasma does not appear in the kinematic regime while magnetic back reaction is negligibly small.

\noindent However, regardless of the anisotropic limit, the fluids driven by a random and isotropic external force eventually become isotropic followed by the strongly frozen magnetic fields if $Pr_M\rightarrow \infty$ ($\eta \rightarrow 0$). Even for a unit magnetic prandtl number ($Pr_M=\nu/\eta=1$), where magnetic fields are not so strongly frozen, the large scale fields driven by the random isotropic force tend to remain independent of direction if a background magnetic field `$\mathbf{b}_{ext}$' is not very strong \citep{Cho et al 2009}. Moreover the growing anisotropy in smaller scales cannot decisively affect the energy spectrum due to the small eddy turnover time. As long as the isotropic force continuously drives the system, especially large scales, the theoretical fluid model that assumes an isotropic system is valid except the case of microscale instability or very strong `$\mathbf{b}_{ext}$'. In addition there is an interesting report that the magnetic energy spectrum around the core of cluster is possibly Kolmogorov's spectrum $\sim k^{-5/3}$ \citep{Kolmogorov 1941} which assumes an isotropic system \citep{Vogt and Ensslin 2003}. Now we can focus our interest on the isotropic properties in ICM of large magnetic prandtl number: energy spectra.\\

\noindent Dynamo in magnetized plasma with large $Pr_M$, which is not rarely observed in space, occurs easily and has its peculiar properties. As the excited magnetic fields (energy) are instilled into the damped kinetic eddies, the viscous damping scale $k_{\nu}\sim 1/l_{\nu}$ is extended toward resistivity scale $k_{\eta}\sim 1/l_{\eta}$ ($k_{\nu}\ll k_{\eta}$). Then, these two coupled velocity and magnetic field generate specific power spectra which are different from the typical spectrum of Kolmogorov's incompressible fluid or Burger's compressible fluid. There have been works on the small scale dynamo for large $Pr_M$ plasma \citep{Schekochihin et al 2002, Cho et al 2003, Schekochihin et al 2004, Yousef et al 2007} in addition to the several analytic methods \citep{Batchelor 1950, Kazantsev 1968, Kulsrud and Anderson 1992, Schober et al 2012, Bovino et al 2013}. To solve the magnetic induction equation MHD theories based on Kazantsev's work \citep{Kazantsev 1968} assume the second order velocity field correlation $\langle \mathbf{v}\cdot \mathbf{v} \rangle\sim (\delta_{ij}-\frac{r_ir_j}{r^2})T_N(r)+\frac{r_ir_j}{r^2}T_L(r)$. However to explain the formation of this second order velocity field correlation, i.e., kinetic energy spectrum, we need to solve the coupled momentum and magnetic induction equation simultaneously instead of making an assumption of the kinetic energy spectrum in advance. Here we use an Eddy Damped Quasi Normal Markovianization method EDQNM, \citep{Kraichnan and Nagarajan 1967, McComb 1990, Park 2013} and dimensional approach in a limited way. We simplified the resultant simultaneous differential equations and found out the solutions for kinetic and magnetic energy spectrum.

\noindent In chapter 2, we briefly introduce simulation tool and analytic method used in this paper. Simulational and analytic results are introduced in chapter 3. And in the final chapter we discuss about the results, their physical meanings, and future topics. Detailed analytic calculations are discussed in the appendix.

\section{Numerical and analytic method}
We have solved the incompressible MHD equations using a pseudo-spectral code with a periodic box of size `$(2\pi)^3$':
\begin{eqnarray}
\frac{\partial \mathbf{v}}{\partial t}&=&-\mathbf{v}\cdot \nabla \mathbf{v} -\nabla P + (\nabla \times \mathbf{B})\times \mathbf{B} + \nu\nabla^2\mathbf{v}+\mathbf{f},\label{Momentum Eq for the code}\\
\frac{\partial \mathbf{B}}{\partial t}&=&\nabla\times(\mathbf{v}\times \mathbf{B})+\mathrm{\mathbf{b}_{ext}},
\label{Magnetic induction Eq for the code}
\end{eqnarray}
where `$\mathbf{f}$' is a random mechanical force driving a system at $k\sim2-3$ in fourier space, and `$\mathrm{\mathbf{b}_{ext}}$' is a weak background (guide) magnetic field\footnote{`$\mathbf{b}_{ext}$' is not indispensable for the growth of magnetic fields if there is a substituting seed magnetic field. The turbulence by cosmological shocks can amplify the weak seed field of any origin \citep{Ryu et al 2008}. It was pointed out that subsonic turbulence could develop with a very weak seed magnetic field \citep{Ryu et al 2012}. Moreover turbulence can amplify a localized seed magnetic field \citep{Brandenburg 2001, Cho and Yoo 2012}. Cho \citep{Cho 2014} investigated the origin of seed magnetic fields and insisted that the origin of the seed field should be more like the localized seed magnetic fields ejected from the astrophysical bodies} ($b_{ext}=0.0001$) covering the whole magnetic eddy scales. Here, `$\mathbf{B}$', magnetic field divided by `$(4\pi\rho)^{1/2}$', has the unit of Alfv$\mathrm{\acute{e}}$n velocity, and `$\mathbf{v}$' is `rms velocity', and `t' has the unit of large scale eddy turnover time `$L/v$'. For example, if `$L$' of a cluster is $\sim 400$ kpc and `$v$' is $\sim 400$ km/s, then `$L/v$' is $\sim 10^9$ year. The time `$t$' has the unit of `$10^9$' year. The system realizes the state of ICM of the high viscous plasma state with ideally frozen magnetic fields. \\

\noindent To compare the effect of hyper diffusivity (incompressible fluid) and physical diffusivity (compressible fluid), we also used $\mathrm{PENCIL\,\, CODE}$ \citep{Brandenburg 2001} with message passing interface(MPI) in a periodic box of spatial volume $(2 \pi)^3$ with mesh size $288^3$. The basic equations solved in the code are,
\begin{eqnarray}
\frac{D \rho}{Dt}&=&-\rho {\bf \nabla} \cdot {\bf u}\\
\frac{D {\bf u}}{Dt}&=&-c_s^2{\bf \nabla} \mathrm{ln}\, \rho + \frac{{\bf J}{\bf \times} {\bf B}}{\rho}+\nu\big({\bf \nabla}^2 {\bf u}+\frac{1}{3}{\bf \nabla} {\bf \nabla} \cdot {\bf u}\big)+{\bf f}\\
\frac{\partial {\bf A}}{\partial t}&=&{\bf u}{\bf \times} {\bf B} -\eta\,{\bf \nabla}{\bf \times}{\bf B}.
\label{MHD equations in the pencil code}
\end{eqnarray}
$\rho$: density; $\bf u$: velocity; $\bf B$: magnetic field; $\bf A$: vector potential; ${\bf J}$: current density;  $D/Dt(=\partial / \partial t + {\bf u} \cdot {\bf \nabla}$): advective derivative; $\eta$: magnetic diffusivity(=$c^2/4\pi \sigma$, $\sigma$: conductivity); $\nu$: kinematic viscosity(=$\mu/\rho$, $\mu$: viscosity); $c_s$: sound speed. Velocity is expressed in units of $c_s$, and magnetic fields in units of $(\rho_0\,\mu_0)^{1/2}c_s$($[B]=\sqrt{\rho_0\,[\mu_0]}[v]$ from $E_M\sim B^2/\mu_0$ and $E_{kin}\sim \rho_0 v^2$). $\mu_0$ is magnetic permeability and $\rho_0$ is the initial density. Note that $\rho_0\sim \rho$ in the weakly compressible simulations. These constants $c_s$, $\mu_0$, and $\rho_0$ are set to be `1'. ${\bf f}(x,t)$ is represented by $N\,{\bf f}_0(t)\, exp\,[i\,{\bf k}_f(t)\cdot {\bf x}+i\phi(t)]$($N$: normalization factor, ${\bf f}_0$: forcing magnitude, ${\bf k}_f(t)$: forcing wave number\footnote{Pencil code selects one of 350 vectors in k$_f$ vector set at each time step. $f_0$ is $0.08$ and injection scale $|k_f|$ is $\sim 1-2$}. The variables are also independent of a unit system. However, instead of $\mathbf{b}_{ext}$, $\mathrm{PENCIL\,\,CODE}$ gives a system initial seed magnetic field in small scales. This seed field disappears in a few simulation time steps. $\mathrm{PENCIL\,\,CODE}$ and the other code \citep{Cho et al 2003} are not the same in various ways, but we will see they produce practically the same energy spectra.


\section{Results}
\subsection{Simulation results}
\begin{figure}
\centering{
  {
   \subfigure[]{
     \includegraphics[width=8cm]{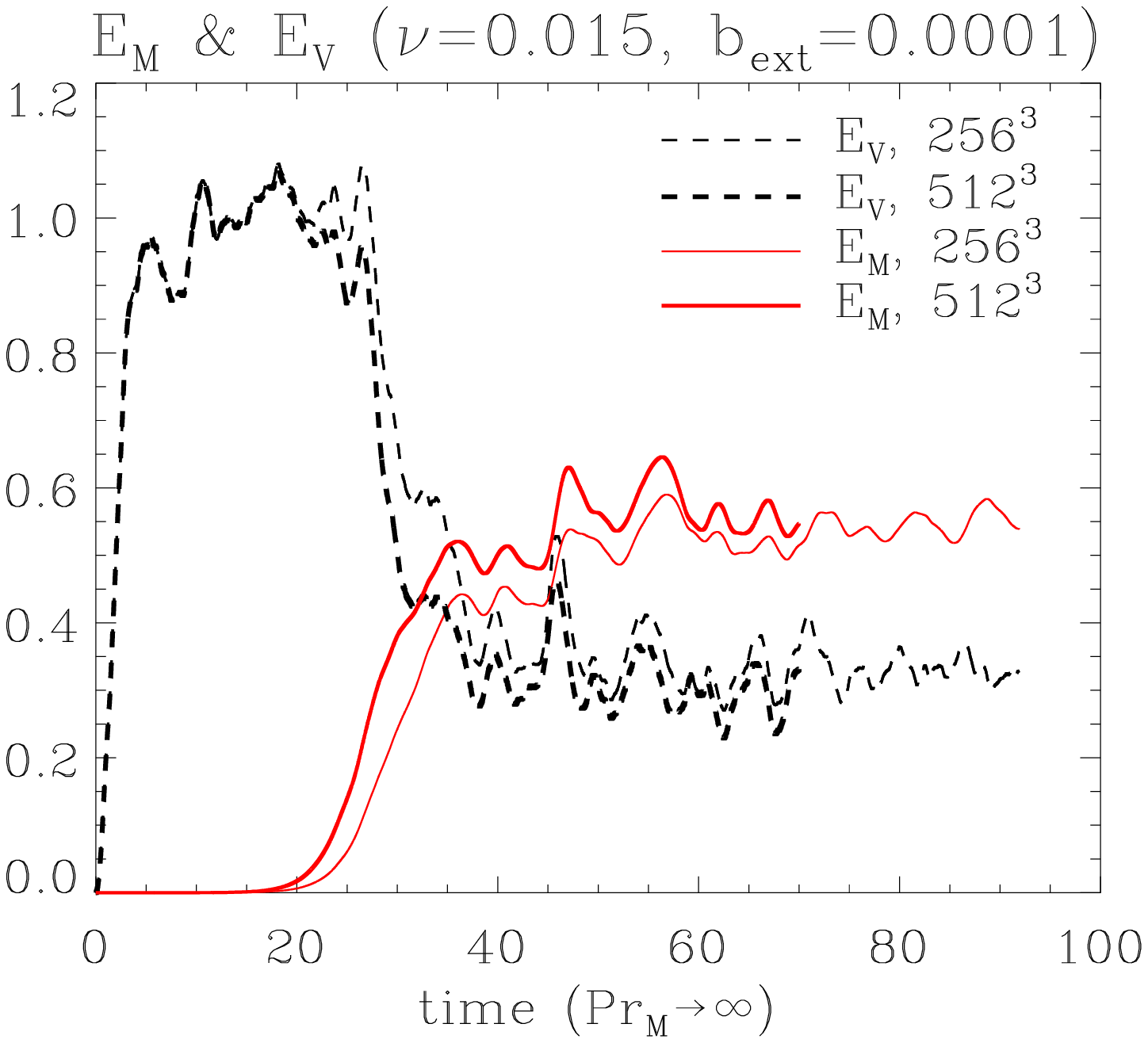}
     \label{f1}
   }\hspace{-10mm}
   \subfigure[]{
     \includegraphics[width=8cm]{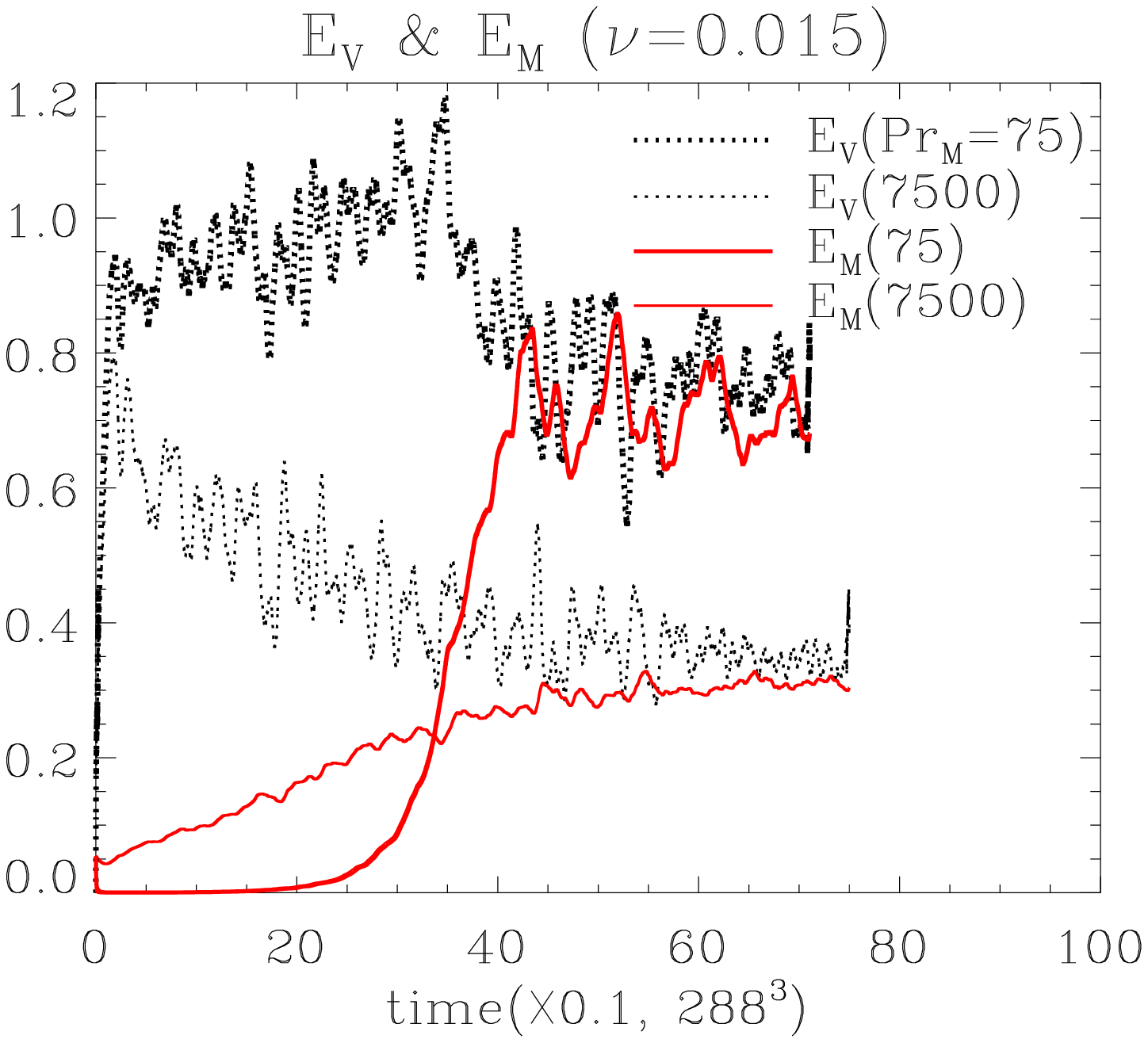}
     \label{f2}
     }
   }
\caption{(a) Normalized $E_V(t)$ and $E_M(t)$ for the incompressible fluid ($Pr_M\rightarrow \infty$, resolution $256^3$ and $512^3$). (b) Normalized $E_V(t)$ and $E_M(t)$ for the compressible fluid ($Pr_M=75$ and $Pr_M=7500$, resolution $288^3$). Time scale is contracted by 10\% ($\times 0.1$).}
}
\end{figure}

Fig.\ref{f1} shows the normalized kinetic energy $E_V(t)$ (black dashed line) and magnetic energy $E_M(t)$ (red solid line) of incompressible MHD fluid ($\nu=0.015$, $\eta\rightarrow 0$, $Pr_M\rightarrow \infty$, $\beta=8\pi P/B^2\rightarrow \infty$) with resolution of $256^3$ (thinner line) and $512^3$ (thicker line). Reynolds number $Re$ of both cases are $\sim 42$, and their magnetic Reynolds number $Re_M$ are actually infinity. When the random isotropic forcing begins to drive the system, $E_V(t)$ quickly grows, becomes saturated, and keeps the status quo until $E_M(t)$ begins to arise at $t\sim 15-20$. As $E_M(t)$ grows, the energy transfer from $E_V(t)$ to $E_M(t)$ gets accelerated. For this nonlocal energy transfer, the gap between kinetic and magnetic energy is an important factor. However, as $\mathbf{B}\cdot \nabla \mathbf{v}$ in Eq.(\ref{Magnetic induction Eq for the code}) implies, the geometrical constraint between $\mathbf{v}$ and $\mathbf{B}$ also plays a role of determinant. If magnetic field is parallel to the gradient of velocity field, kinetic energy is transferred to magnetic eddies. If not, magnetic field frozen to the plasma fluid moves freely along the fluid or gets annihilated. Around the onset position at $t\sim 15-20$, structural change between $E_V$ and $E_M$ seems to get accelerated and completed at the saturation. Also the plot shows the onset position of $E_M(t)$ is proportional to the resolution. Higher resolution without resistivity makes more space where the forward cascaded $E_M(t)$ can stay not being dissipated much. This leads to the imbalanced energy distribution like the relatively smaller amount of magnetic energy in large scales and more amount of magnetic energy in small scales. This makes the gap between $E_V$ and $E_M$ in large scales, especially around the injection scale, increase so that the nonlocal energy transfer become accelerated. The saturated magnetic energy $E_{M, sat}$ grows with the increase of resolution, but the saturated kinetic energy $E_{V, sat}$ decreases. This is the special feature of $\eta\sim 0$ with a fixed high viscosity.\\

\noindent Fig.\ref{f2} includes the normalized $E_V(t)$ and $E_M(t)$ of compressible fluids. Their properties are as follows; for thicker line, $Pr_M$=75, $\nu=0.015$, $\eta=2\times 10^{-4}$, $Re\sim 170$, $Re_M\sim 1.3\times 10^4$, $\beta\sim 83$; for the thinner line, $Pr_M$=7500, $\nu=0.015$, $\eta=2\times 10^{-6}$, $Re\sim 120$, $Re_M\sim 8.8\times 10^5$, $\beta\sim 430$. The resolutions in both cases are $288^3$. Since the average of Mach number is at most 0.18, the effect of compressibility is not much. We can infer more $E_V$ can be transferred to $E_M$ with higher $Pr_M$ (lower $\eta$). However, the plot shows the actual $E_{V, sat}$ \& $E_{M, sat}$ are inversely proportional to $Pr_M$. Moreover $E_{V, sat}$ is slightly larger than $E_{M, sat}$, and this tendency is opposite to that of hyper diffusivity (Fig.\ref{f1}). If $\eta$ is small, magnetic energy can migrate into the smaller scales. But since the effect of dissipation grows with the wave number $k^2$, more $E_M$ is dissipated to increase energy gap between $E_V$ and $E_M$. This boosts the nonlocal energy transfer bringing about the accelerated dissipation of energy ($\sim k^2 E(k)$) in small scales. So the saturated energy $E_{V, sat}$, $E_{M, sat}$ are smaller than those of lower $Pr_M$. However, it is not easy to conclude whether they converge on some lower limit as the physical diffusivity $\eta\rightarrow 0$ with the limited resolution and $Pr_M$ at this moment.\\

\begin{figure}
\centering{
  {
   \subfigure[]{
     \includegraphics[width=8cm]{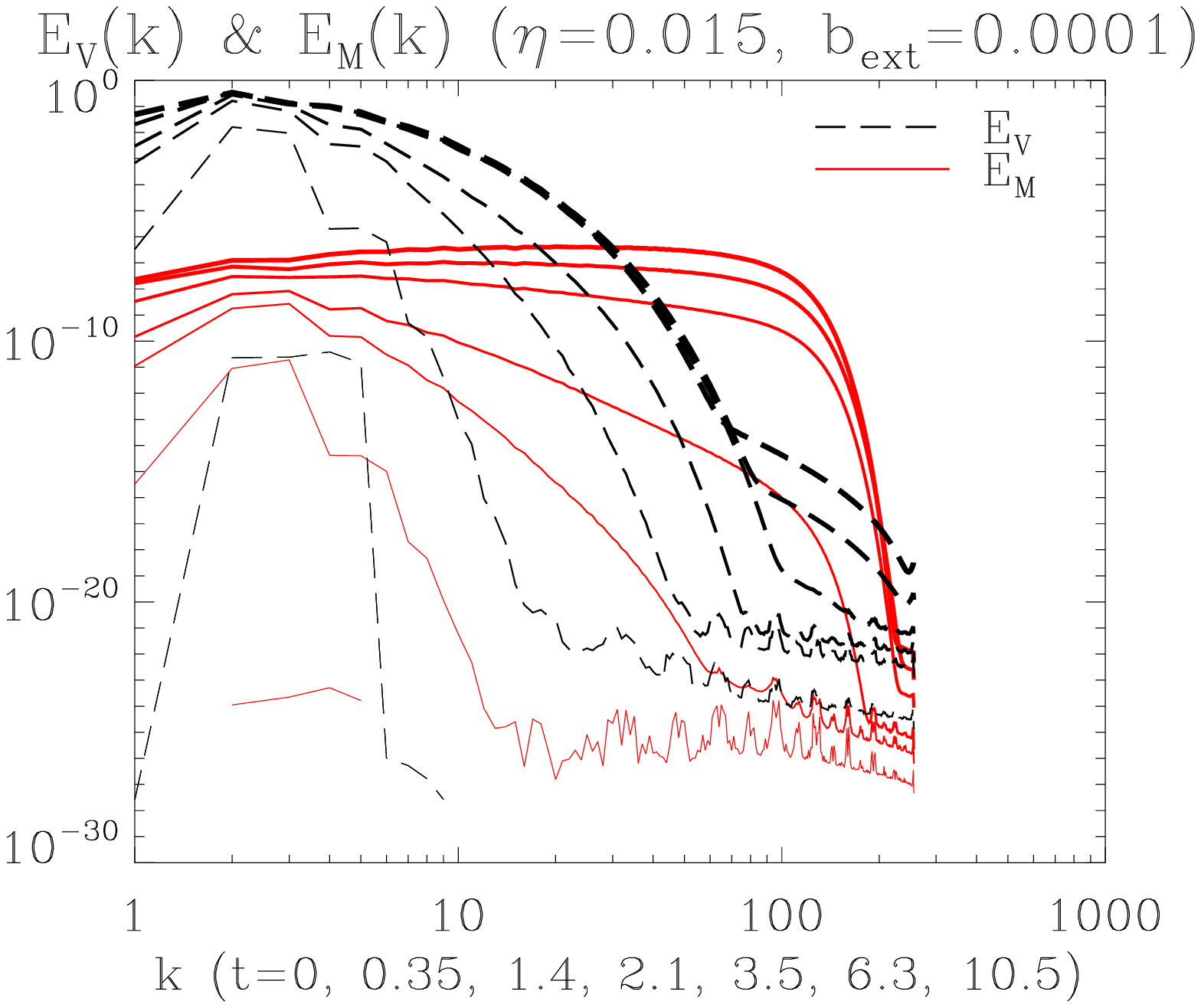}
     \label{f3}
   }\hspace{-10mm}
   \subfigure[]{
     \includegraphics[width=8cm]{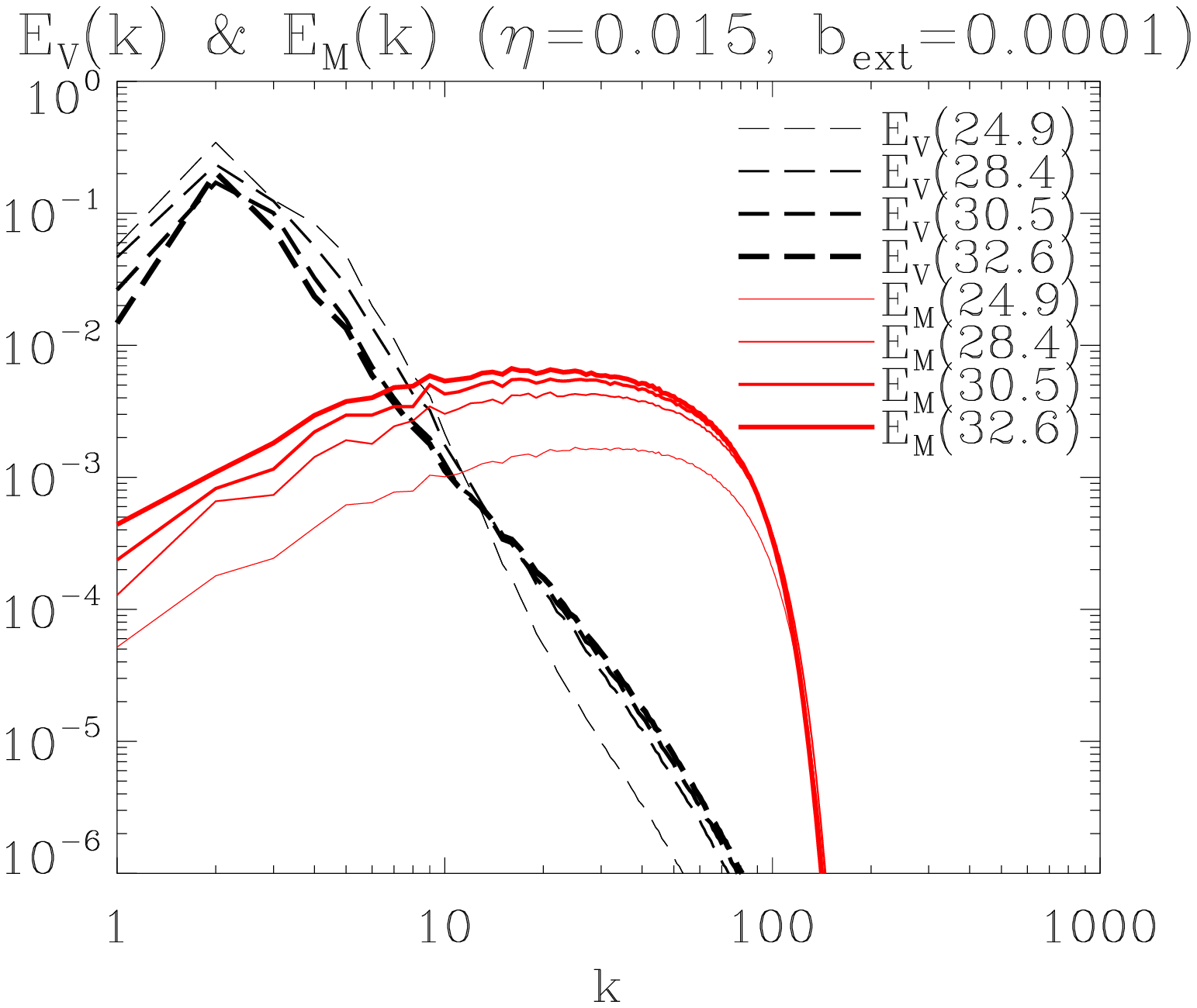}
     \label{f4}
     }
  }
\caption{Energy spectra of incompressible fluid ($Pr_M\rightarrow \infty$) (a) $E_V(k)$ and $E_M(k)$ in the early time regime ($t\leq 10.5$) (b) $E_V(k)$ and $E_M(k)$ in $24.9\leq t \leq 32.6$. As $E_M$ grows, large scale kinetic energy is transferred to magnetic eddies, but kinetic eddies in subviscous scale receive energy from the magnetic eddies.}
}
\end{figure}

\noindent Fig.\ref{f3} shows the evolving energy spectrum $E_V(k)$ and $E_M(k)$ of the incompressible fluid in the early time regime $t\sim 0-10.5$ (from bottom to top). And the plot in Fig.\ref{f4} is a de facto the same one in the range of $t=24.9-32.6$. When the kinetic energy begins to drive the system at $k=2.5$, $E_V$ (black dashed line, $t\sim 0$) grows prior to $E_M$ (red solid line). As the advection term $\mathbf{v}\cdot \nabla \mathbf{v}$ indicates, $E_M$ is not necessary for the local energy transfer in kinetic eddies. However,  without $E_V(k)$ magnetic eddies cannot receive energy from kinetic eddies nor transfer its energy to the neighboring eddies. Only when magnetic fields run into the plasma fluids with the nontrivial $E_V$ of which gradient is in the direction of magnetic fields, dynamo process occurs leading to the growth of $E_M$. Thus, in the very early time regime the evolution of magnetic field looks subsidiary. But around $t\sim 1.4-2.1$ $E_M(k)$ in small scales begins to surpass $E_V(k)$, and gets past $E_V(k)$ which suffers from the viscous dissipation. So most of $E_V$ is located near the injection scale where the viscous damping effect is not so much, but $E_M$ cascades forward and stays in the smaller scales. Kinetic eddies in $k<\sim 15$ lose energy (or magnetic eddies receive energy through $\mathbf{B}\cdot \nabla \mathbf{v}$) whereas smaller scale kinetic eddies ($k>\sim 15$) receive energy through magnetic tension $\mathbf{B}\cdot \nabla \mathbf{B}$. As the fluid motion can amplify the magnetic field through dynamo process, also magnetic fields can increase the kinetic energy through Lorentz force ($\mathbf{J}\times \mathbf{B}$). The magnetic fields press the fluid through magnetic pressure ($-\nabla B^2/2$) and stretch the fluid through magnetic tension. We will see that the unusual $E_V(k)$ spectrum $k^{-3}$ is due to not only the viscous damping but also the interaction with $E_M$.\\

\begin{figure}
\centering{
  {
     \subfigure[]{
     \includegraphics[width=8cm]{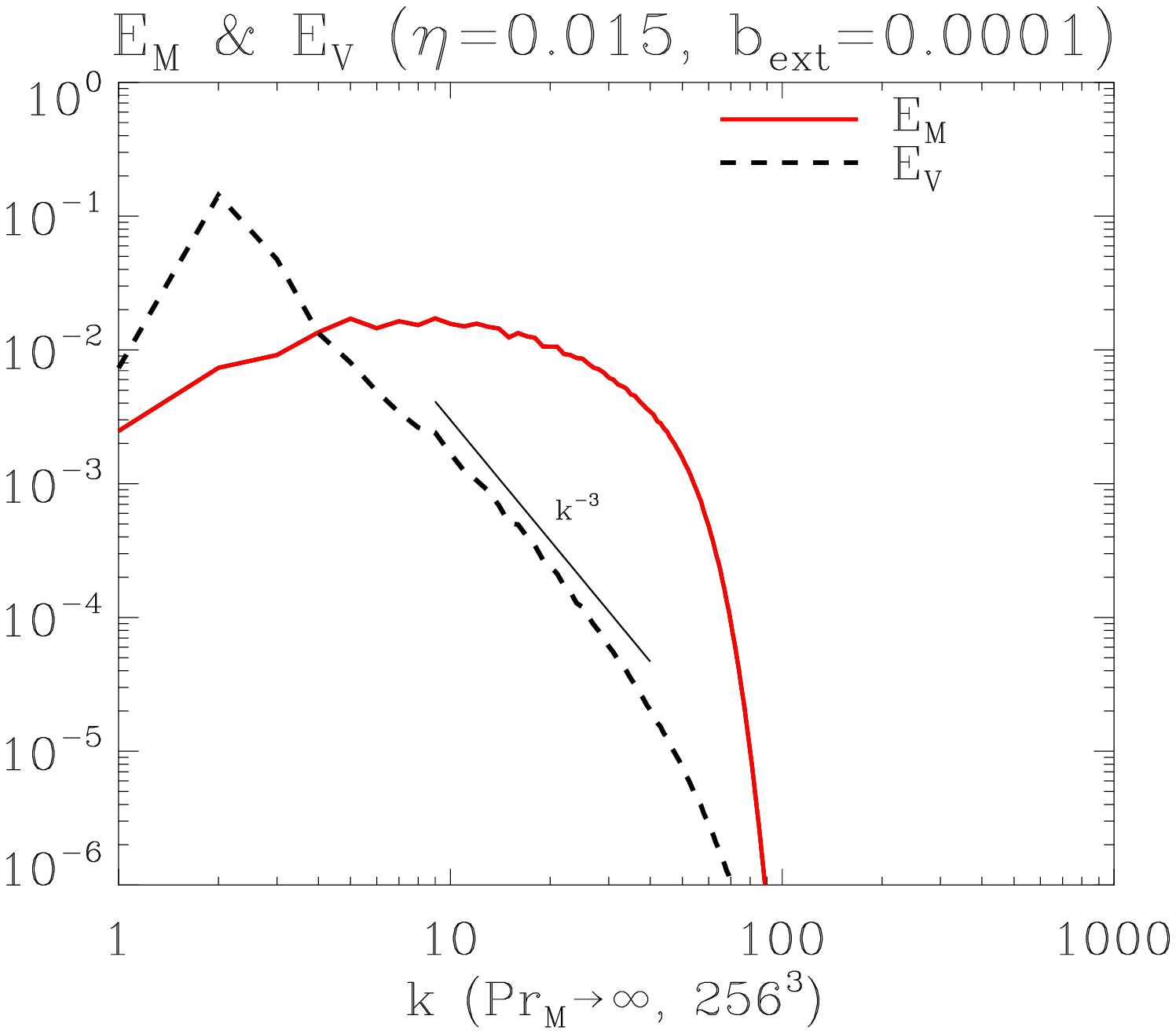}
     \label{f5}
     }\hspace{-10mm}
   \subfigure[]{
     \includegraphics[width=8cm]{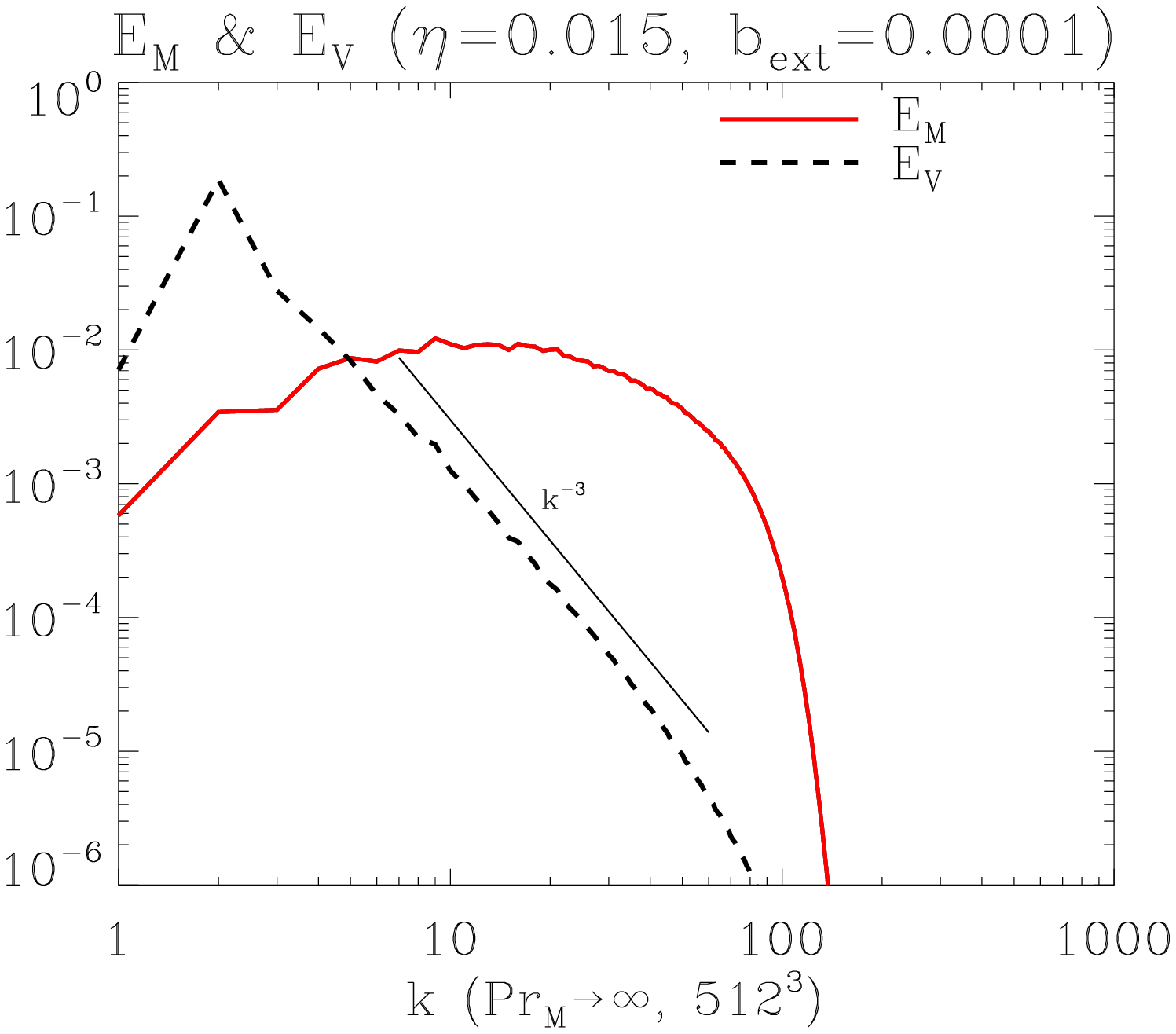}
     \label{f6}
     }\hspace{-10mm}
   \subfigure[]{
     \includegraphics[width=8cm]{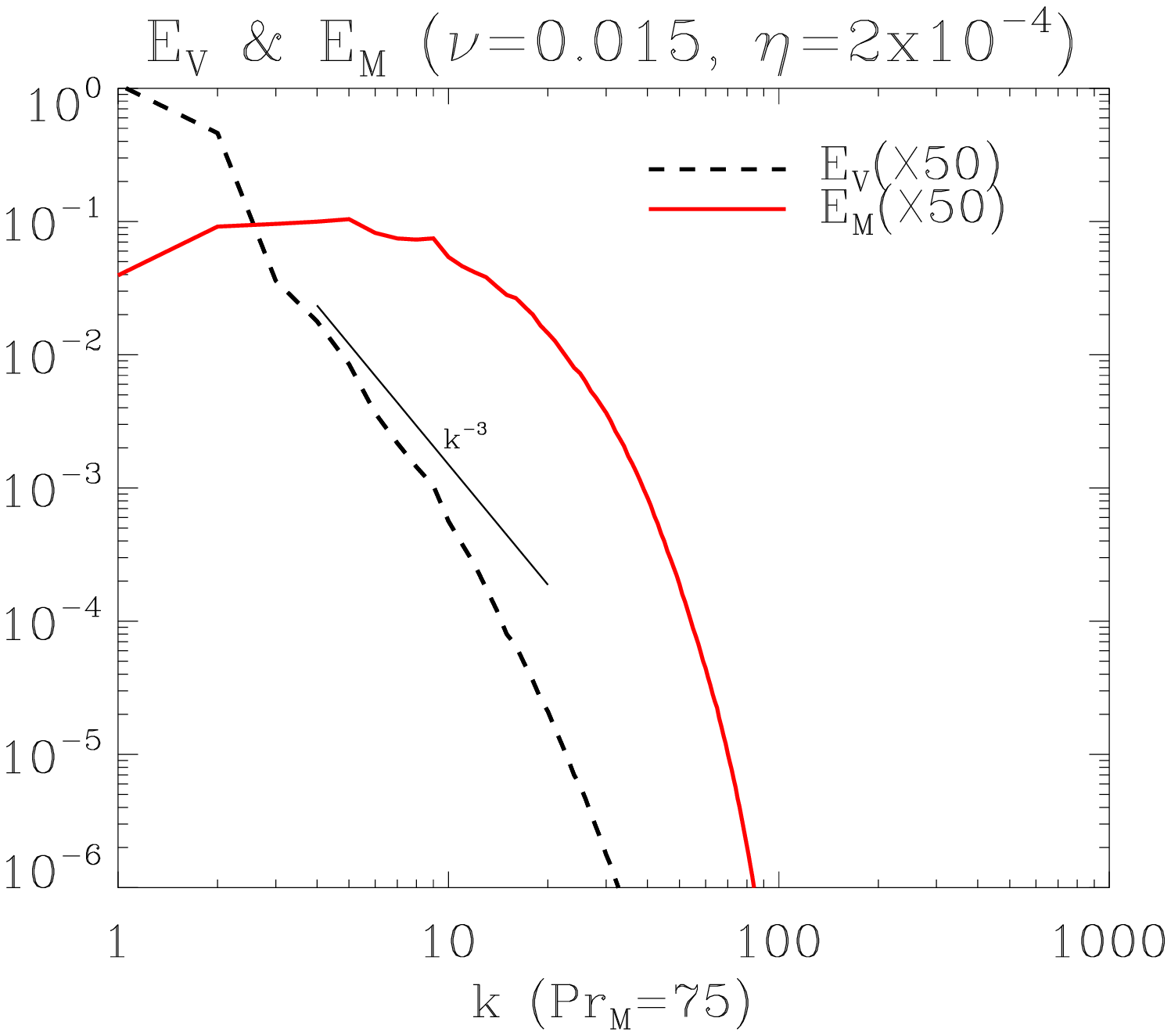}
     \label{f7}
   }\hspace{-10mm}
   \subfigure[]{
     \includegraphics[width=8cm]{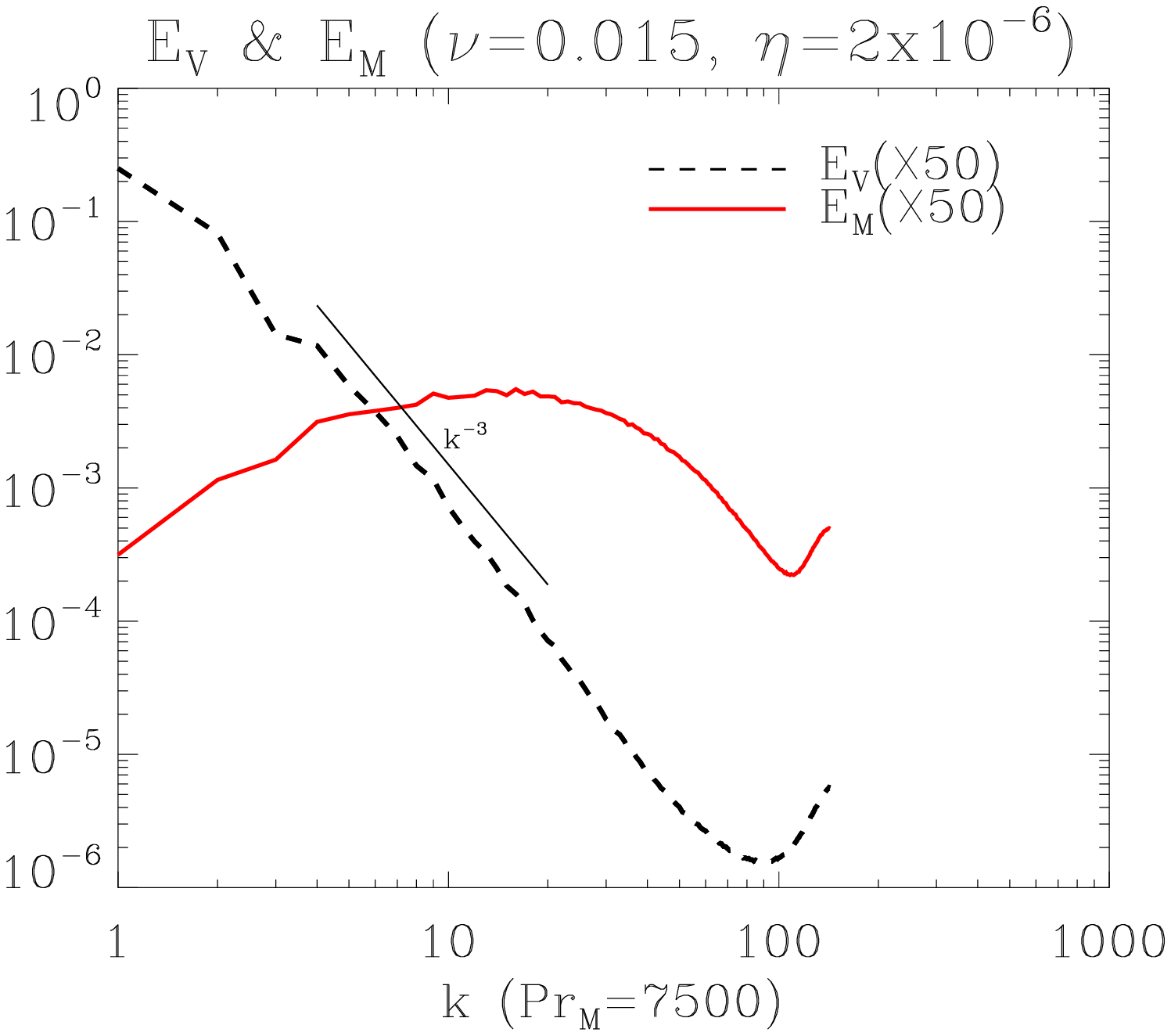}
     \label{f8}
  }
}

\caption{(a), (b) Saturated $E_V(k)$ and $E_M(k)$ with resolution 256$^3$ and 512$^3$ for the incompressible fluid. (c), (d) Saturated $E_V(k)$ and $E_M(k)$ for the compressible fluids with resolution $288^3$. Saturated energy level of $Pr_M=75$ is higher than that of $Pr_M=7500$. Bottle neck effect appears.}
}
\end{figure}

\noindent Fig.\ref{f5}, \ref{f6} include the saturated energy spectra of incompressible fluids with hyper diffusivity. Also the saturated energy spectra of compressible fluids with physical diffusivity are shown in Fig.\ref{f7}, \ref{f8}. These plots show $E_{V,sat}(k)$ eventually converges to $\sim k^{-3}$ if $Pr_M$ is not too small regardless of the different evolving profiles of $E_V(t)$ and $E_M(t)$ in Fig.\ref{f1}, \ref{f2}. So the comparison of these plots will give us some clues to the formation of $E_V(k)$ in small scale range. A quick look shows $E_V(k)$ of $512^3$ has clear spectrum of $k^{-3}$ compared with the kinetic energy spectrum of $256^3$. However, these two simulation sets have the same conditions: weak $\mathbf{b}_{ext}=0.0001$, isotropic random driving force $\mathbf{f}$, injection scale $k_f\sim 2.5$, viscosity $\nu=0.015$, and negligible magnetic diffusivity $\eta$. Also the saturated energy levels shown in Fig.\ref{f1} are not much different. Just the distribution of $E_M$ of $512^3$ in small scale is flatter than that of $256^3$. This implies $E_M$ in smaller scales may be a determinant of $E_V$ spectrum.\\

\noindent On the other hand, for the compressible system in Fig.\ref{f7} and Fig.\ref{f8}, $E_V(k)$ of $Pr_M=7500$ has clearer and longer spectrum of $k^{-3}$ despite its smaller saturated energy level: $E(t)_{sat.,\, Pr_M=75}>E(t)_{sat.,\, Pr_M=7500}$(Fig.\ref{f2}). This means $E_{V, sat.}(k)$ is not so much influenced by the magnitude of $E_M(k)$. However, careful look shows the peak of $E_M$ in Fig.\ref{f8} is located at smaller scale regime than that of Fig.\ref{f7}. Then kinetic eddies in wider range can interact with magnetic energy $E_M(k)$ whose power spectrum is less steep. In other words, Fig.\ref{f6} and Fig.\ref{f8} commonly show the slope of $E_M(k)$ is not so slanted as that of Fig.\ref{f5} and Fig.\ref{f7}.\\

\noindent So if we assume a critical representative spectrum $E_M(k)\sim k^m$ in the subviscous regime, we can infer that steeper spectrum than $k^m$ cannot provide the kinetic eddies with enough energy for $E_V\sim k^{-3}$. However the exact value of `$m$' is not known yet. Just we can guess this index should be negative. The scaling factor `$k^{-1}$', so called an invariant scaling factor \citep{Ruzmaikin et al 1982, Kleeorin et al 1996}, is drawn together in next figure for reference. But we will not discuss this concept further in this paper.\\

\begin{figure}
\centering{
  {
   \subfigure[]{
     \includegraphics[width=8cm]{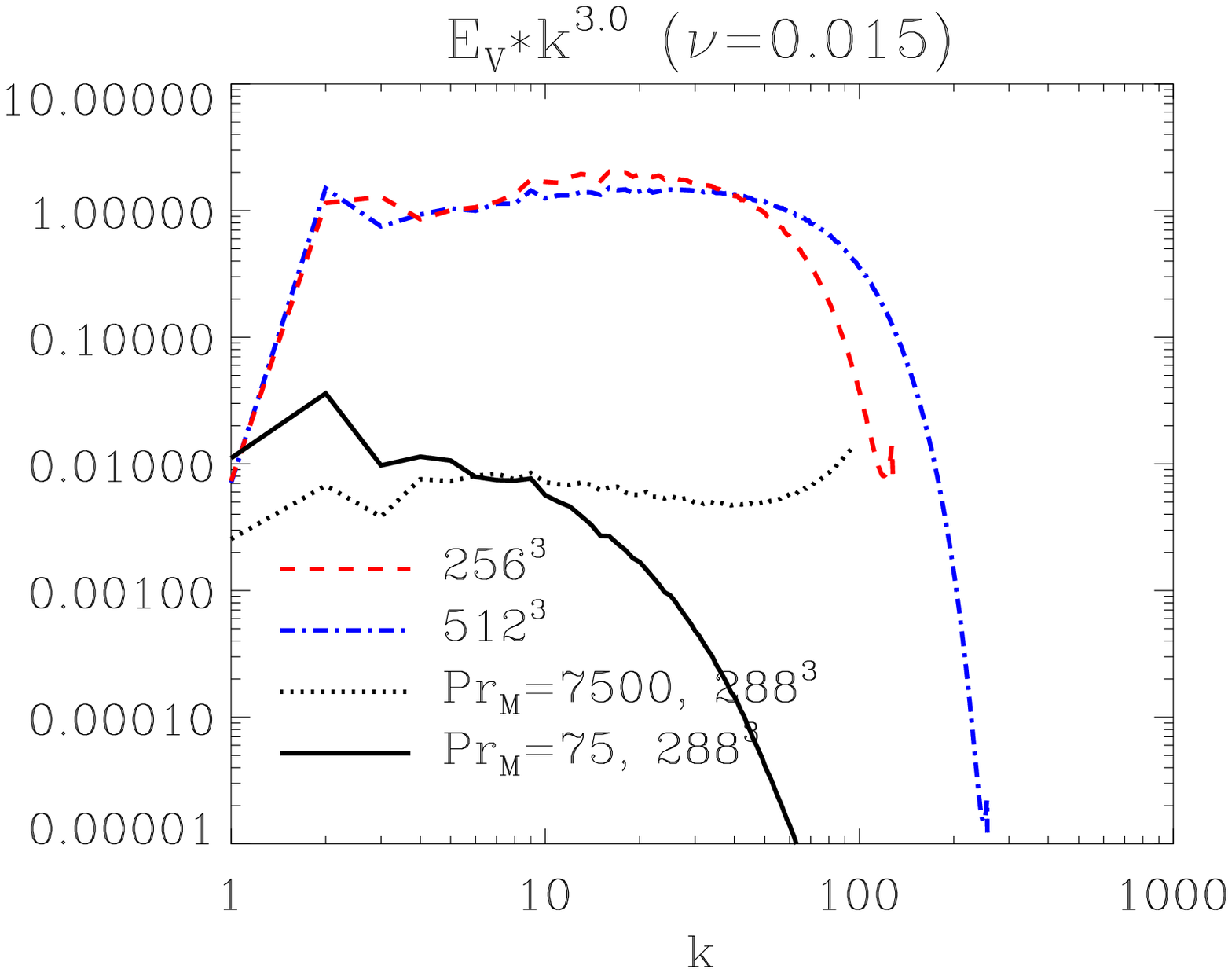}
     \label{f9}
   }\hspace{-10mm}
   \subfigure[]{
     \includegraphics[width=8cm]{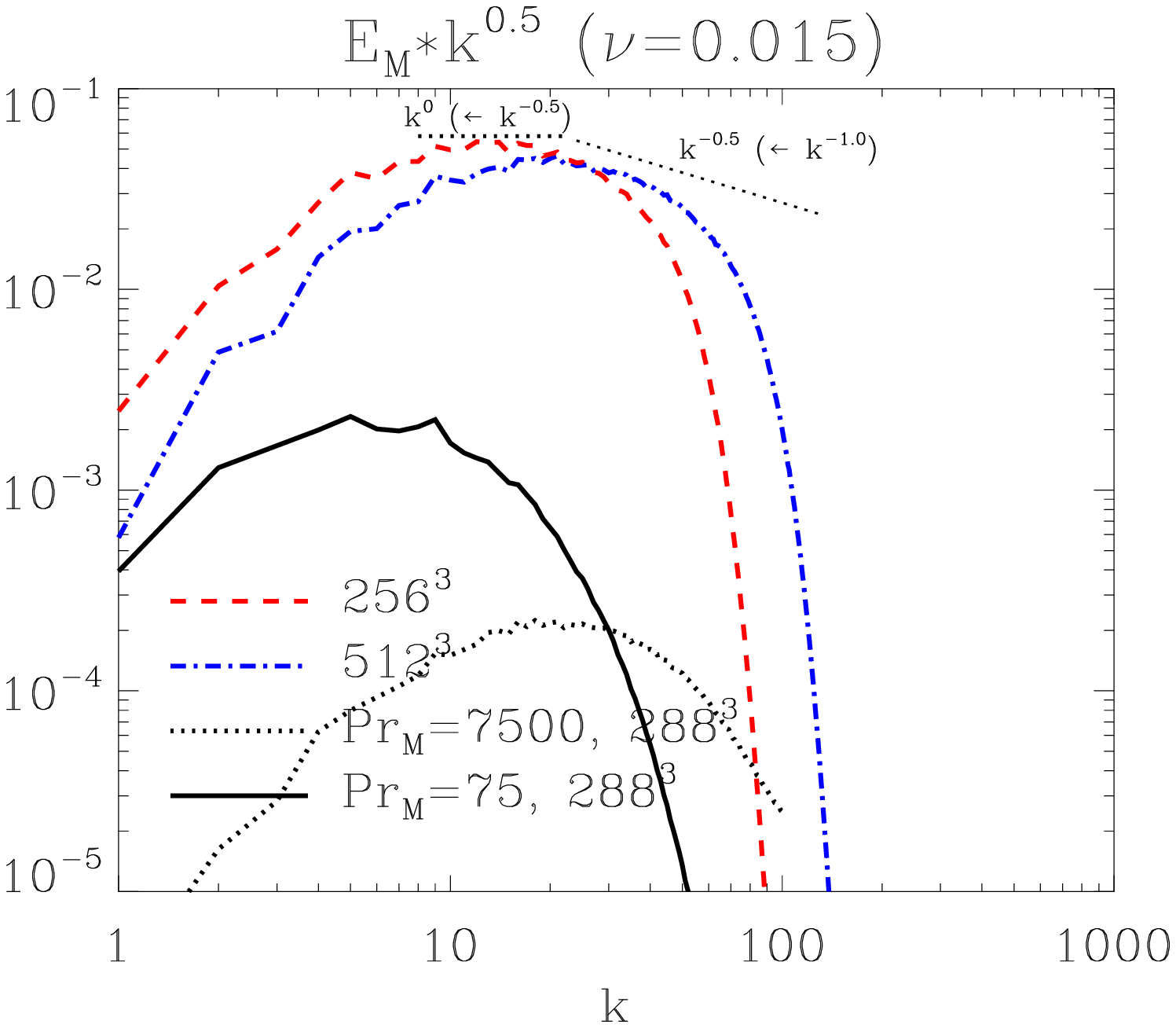}
     \label{f10}
     }
  }
\caption{(a) Compensated energy spectrum $k^3E_V(k)$. (b) $k^{0.5}E_M(k)$. The flat reference line $k^0$ means $k^{-1/2}$ in $E_M(k)$, and slanted line $k^{-0.5}$ is the scaling invariant line $k^{-1}$.}
}
\end{figure}

\noindent Fig.\ref{f9}, \ref{f10} include the compensated $k^3E_V$ and $k^{1/2}E_M$. $E_V(k)$ of high $Pr_M$ has clear spectrum of $k^{-3}$ in the subviscous scale. But for the spectrum of $E_M$ in Fig.\ref{f10}, it is ambiguous to pinpoint any scaling factor. Instead, we choose `$k^{-1/2}$', i.e., $m=-1/2$, an approximate median value of the previous results \citep{Cho et al 2003, Lazarian et al 2004, Schekochihin et al 2004}. We will see analytic method derives quite exact $E_V(k)$ with this scaling factor $k^{-1/2}$. A reference line `$k^{-1}$' for the scaling invariant factor is drawn together.\\

\noindent Fig.\ref{f11} shows $E_M/E_V$ for fig.\ref{f5}-\ref{f8}. The ratio of hyper diffusivity is larger than that of physical diffusivity, which is consistent. As mentioned, with the negligible diffusivity the ratio proportionally depends on the resolution because of the larger $E_M$. In contrast, for the case of physical diffusivity (black line), the dissipation effect growing with wave number ($\sim k^2$) dissipates $E_M$ more efficiently until the energy states are balanced to be in the state of equipartition.\\

\subsection{Analytic methods \& results}
\subsubsection{Eddy Damped Quasi Normal Markovianization}
\noindent We start from the momentum (Eq.\ref{Momentum Eq for the code}) and magnetic induction equation (Eq.\ref{Magnetic induction Eq for the code}). Taking divergence of the momentum equation we can replace pressure by convection and magnetic tension (Leslie and Leith 1975, Yoshizawa 2011)\footnote{`$v(k,t)$' and `$B(k,t)$' depend on time `$t$' and wavenumber `$k$', but `t' will be omitted for simplicity.}:
\begin{eqnarray}
\frac{\partial v_i(k,t)}{\partial t}&=&\sum_{p+r=k}M_{iqm}(k)[v_q(p,t)v_m(r,t)-B_q(p,t)B_m(r,t)]+\nu \nabla^2 v_i(k,t),\label{momentuem equation in Fourier space}\nonumber\\
\\
\frac{\partial B_i(k,t)}{\partial t}&=&\sum_{p+r=k}M^B_{iqm}(k)v_q(p,t)B_m(r,t),\label{magnetic induction equation in Fourier space}
\end{eqnarray}
with the definition of algebraic multipliers
\begin{eqnarray}
M_{iqm}(k)&=&-\frac{i}{2}\big(k_m\delta_{iq}+k_q\delta_{im}-\frac{2k_ik_qk_m}{k^2}\big),\nonumber\\
M^B_{iqm}(k)&=&i(k_m\delta_{iq}-k_q\delta_{im}).
\label{definition of M}
\end{eqnarray}
Then we can get the evolving second order correlation equations of $\langle v_i(k) v_i(-k)\rangle$ and $\langle B_i(k) B_i(-k)\rangle$:
\begin{eqnarray}
\big(\frac{\partial}{\partial t}+2\nu k^2 \big)\langle v_i(k)v_i(-k)\rangle=\sum_{p+r=k}M_{iqm}(k)[\overbrace{\langle v_q(p)v_m(r)v_i(-k)\rangle}^{A1}-\overbrace{\langle v_q(-p)v_m(-r)v_i(k)\rangle}^{A2}\nonumber\\
-\overbrace{\langle B_q(p)B_m(r)v_i(-k)\rangle}^{A3}+\overbrace{\langle B_q(-p)B_m(-r)v_i(+k)\rangle}^{A4}],
\label{momentum correlation equation in Fourier space}
\end{eqnarray}
\begin{eqnarray}
\frac{\partial}{\partial t}\langle B_i(k)B_i(-k)\rangle=\sum_{p+r=k}M^B_{iqm}(k)[ \overbrace{\langle v_q(p)B_m(r)B_i(-k)\rangle}^{B1}-\overbrace{\langle v_q(-p)B_m(-r)B_i(k)\rangle}^{B2}].
\label{magnetic field correlation equation in Fourier space}
\end{eqnarray}
The third order correlation, `$A1, A2,...,B2$', is called a transport function. These triple correlations play a role of transfer and dissipation of energy to decide the field profiles of the system. If the field is helical, `$\alpha$' coefficient ($\sim \langle \mathbf{j}\cdot \mathbf{b}\rangle-\langle \mathbf{v}\cdot \mathbf{\omega}\rangle$) for the inverse cascade of magnetic energy can be derived. Also other terms for the forward cascade of energies are derived \citep{Krause and Radler 1980, Moffatt 1978, Park and Blackman 2012a, Park and Blackman 2012b, Pouquet et al 1976}. However the helical field is not assumed in our system, which means `$\alpha$' coefficient or the inverse cascade of magnetic energy is excluded. We use a quasi normalization approximation, sort of an iterative method, to find the transport function (appendix).\\

\noindent The formal representations of $E_V(k)$ and $E_M(k)$ are as follows:
\begin{eqnarray}
\frac{\partial E_V(k)}{\partial t}&=&+\frac{1}{2}\int dp\,dr\,\Theta^{\nu\nu\nu}_{kpr}(t)\frac{k^3}{p\,r}
(1-2y^2z^2-xyz)E_V(p)E_V(r)\nonumber\\
&&-\int dp\,dr\,\Theta^{\nu\nu\nu}_{kpr}(t)\frac{p^2}{r}(xy+z^3)E_V(r)E_V(k)-2\nu k^2 E_V(k)\nonumber\\
&&+\frac{1}{2}\int dp\,dr\,\Theta^{\nu\eta\eta}_{kpr}(t)\frac{k^3}{p\,r}(1-2y^2z^2-xyz)E_M(p)E_M(r)\nonumber\\
&&+\int dp\,dr\,\Theta^{\nu\eta\eta}_{kpr}(t)\frac{p^2}{r}(y^2z-z)E_M(r)E_V(k),\nonumber\\
\label{EDQNMEv(t)}
\end{eqnarray}
and
\begin{eqnarray}
\frac{\partial E_M(k)}{\partial t}&=&-\int dp\,dr\,\Theta^{\eta\eta\nu}_{krp}(t)\frac{p^2}{r}
z(1-x^2)E_M(r)E_M(k)\nonumber\\
&&-\int dp\,dr\,\Theta^{\eta\eta\nu}_{krp}(t)\frac{r^2}{p}
(y+xz)E_V(p)E_M(k)\nonumber\\
&&+\int dp\,dr\,\Theta^{\eta\eta\nu}_{krp}(t)\frac{k^3}{p\,r}
(1+xyz)E_V(p)E_M(r).\nonumber\\
\label{EDQNMEm(t)}
\end{eqnarray}
The integral variables `$\mathrm{p}$', `$\mathrm{r}$', i.e., wavenumbers, are constrained by the relation of $p+r=k$. `$x$', `$y$', and `$z$' are cosines of the angles formed by three vectors `$\mathbf{k}$', `$\mathbf{p}$', and `$\mathbf{r}$'. (Fig.\ref{f11}). Algebraically `$k$', `$p$', and `$r$' should satisfy a condition like (Leslie and Leith 1975):
\begin{eqnarray}
|k-r|\leq p \leq k+r.
\label{three vectors summation condition}
\end{eqnarray}
To derive analytically solvable equations from Eq.(\ref{EDQNMEv(t)}), (\ref{EDQNMEm(t)}), we need to simplify these two equations considering the interaction between `$k$' and its close wave number. So, we take account of only two cases: large $p$ ($k\sim p\gg r$) and large $r$ ($k\sim r\gg p$). In principle `$k/2\sim p\sim r$' should also be included. But since the interaction between the close wave vectors, local energy transfer, is dominant in the nonhelical small scale dynamo, the latter case can be ignored.\\

\begin{figure}
\centering{
  {
   \subfigure[$\mathbf{p}+\mathbf{r}+\mathbf{k}$]{
     \includegraphics[width=5.6cm]{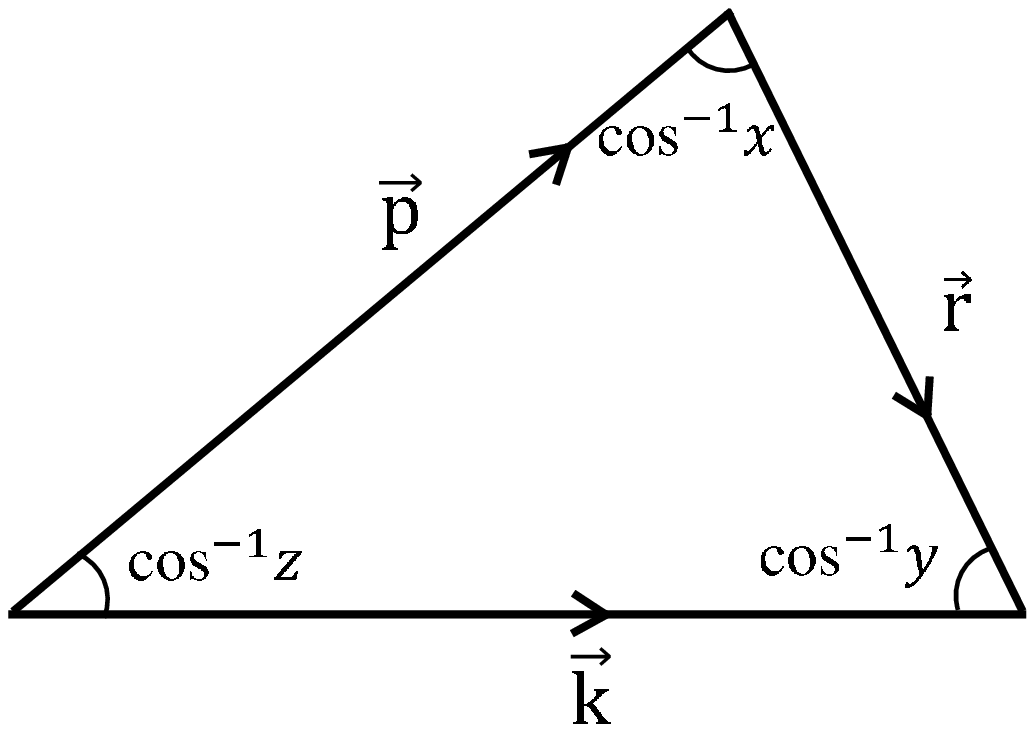}
     \label{f12}
   }\quad
   \subfigure[$k\sim p \gg r$, $k\sim r \gg p$]{
     \includegraphics[width=6.5cm]{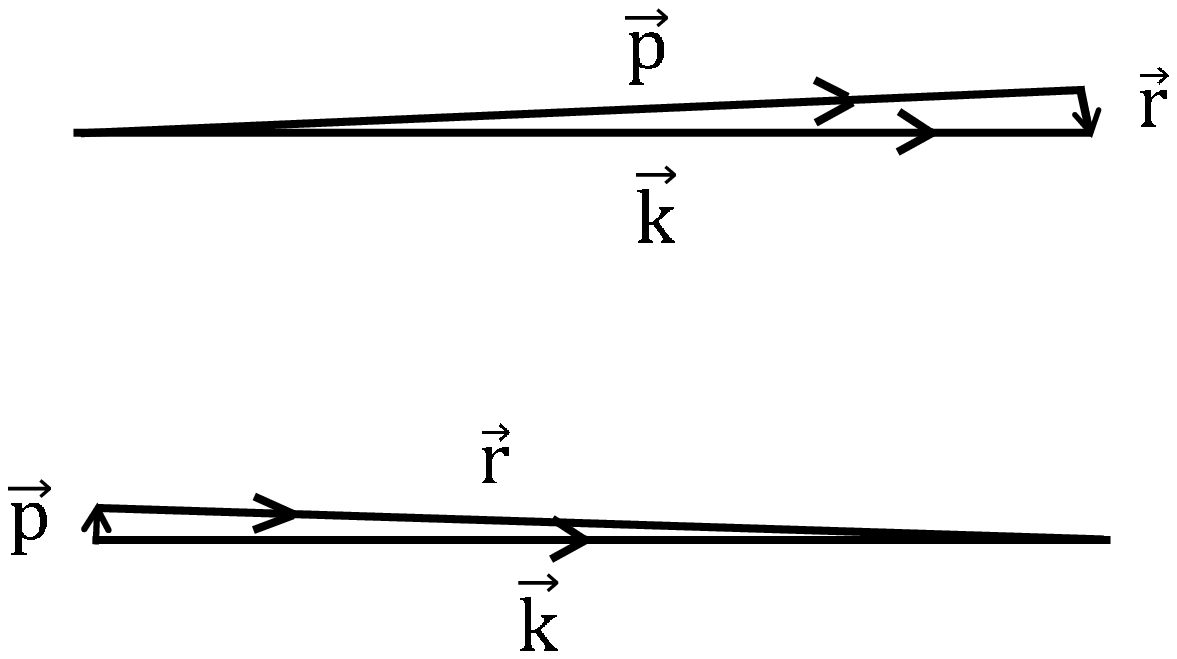}
     \label{f13}}
  }
\caption{(a) The summation of three wave numbers should satisfy the condition $\mathbf{p}+\mathbf{r}+\mathbf{k}=0$ (b) Upper triangle is the case of large $\mathbf{p}$ (small $\mathbf{r}$): $k\sim p \gg r$, $x\sim y$, $z\sim 1$, and lower one is for large $\mathbf{r}$ (small $\mathbf{p}$): $k\sim r\gg p$, $x\sim z$, $y\sim 1$.}
}
\end{figure}

\noindent With the assumption of $E_V(k)\sim k^v$ and $E_M(k)\sim k^m$, we simplify the first term in Eq.(\ref{EDQNMEm(t)}) like below:\\
(i) Large $p$ (small $r$, i.e., $k\sim p\gg r$, $\eta\sim0$)\\
Eddy damping function $\Theta_{krp}^{\eta\eta\nu}(k,t)$ is approximately
\begin{eqnarray}
\Theta_{krp}^{\eta\eta\nu}(k,t)=\frac{1-e^{-[\nu p^2+\eta (k^2+r^2)+\mu_{kpr}]t}}{\nu p^2+\eta (k^2+r^2)+\mu_{kpr}}\bigg|_{\eta=0}\sim\frac{1}{\nu p^2},\qquad(t\rightarrow \infty).
\label{simplifying coefficient large p1}
\end{eqnarray}
Also as Fig.\ref{f11}, \ref{f12} show, we can use the relations $x\sim y\sim 0$, $z\sim 1$, and $r=k-p$. Then,
\begin{eqnarray}
-\int^k dp \frac{1}{\nu p^2}\frac{p^2}{r}(1-x^2)\,(k-p)^mE_M(k)\sim -\frac{1}{\nu} k^mE_M(k).
\label{simplifying coefficient large p2}
\end{eqnarray}
(ii) Small $p$ (large $r$, i.e., $k\sim r\gg p$, $x\sim z\sim 0$, $y\sim 1$)\\
In this case only the triad relaxation time $\mu_{kpr}$ is left. We assume that it is independent of time, so we can write $\mu_{kpr}$ like
\begin{eqnarray}
\Theta_{krp}^{\eta\eta\nu}(k,t)\sim\frac{1}{\nu p^2+\mu_{kpr}}\sim\frac{1}{\mu_{kpr}}.
\label{simplifying coefficient large r1}
\end{eqnarray}
Then,
\begin{eqnarray}
\sim-\int^k dr \frac{1}{\mu_{kpr}}(k^2r^{m-1}-2kr^m+r^{m+1})x(1-x^2)E_M(k)\sim 0.
\label{simplifying coefficient large r2}
\end{eqnarray}
The results, Eq.(\ref{simplifying coefficient large p2}), (\ref{simplifying coefficient large r2}) represent the first term in `$D$' in Eq.(\ref{Em(t)}). The other terms can be found in a similar way.\\

\noindent Then, the coupled equations of $E_V(k)$ and $E_M(k)$ are
\begin{eqnarray}
\frac{\partial E_V(k)}{\partial t}=-\underbrace{\big(a_1k^{v}+2\nu k^2\big)}_{A}E_V(k)
+\underbrace{\big(b_1k^m-b_2k^v\big)}_{B}E_M(k),
\label{Ev(t)}
\end{eqnarray}
\begin{eqnarray}
\frac{\partial E_M(k)}{\partial t}=\underbrace{\big(c_1k^m+c_2k^{m+2}\big)}_CE_V(k)-\underbrace{\big(d_1k^m+d_2k^{v+2}\big)}_DE_M(k).
\label{Em(t)}
\end{eqnarray}
The coefficients `$a_i$', `$b_i$', `$c_i$', and `$d_i$' are independent of `$k$', and assumed to be independent of time for simplicity. The matrix form of these simultaneous differential equations is simply
\begin{eqnarray}
\left[
  \begin{array}{c}
    E'_V(k) \\
    E'_M(k) \\
  \end{array}
\right] = \underbrace{\left[
            \begin{array}{cc}
              -A & B \\
              C & -D \\
            \end{array}
          \right]}_\mathbf{\mathcal{M}}\left[
  \begin{array}{c}
    E_V(k) \\
    E_M(k) \\
  \end{array}
\right].
\label{Ev(t)_Em(t)_matrix}
\end{eqnarray}
Eq.(\ref{Ev(t)_Em(t)_matrix}) can be solved diagonalizing the matrix `$\mathcal{M}$'. For this, the bases $E_V(k)$ and $E_M(k)$ need to be transformed using a transition matrix `$\mathcal{P}$' which is composed of eigenvectors of `$\mathcal{M}$':
\begin{eqnarray}
\left[
  \begin{array}{c}
    E_V(k) \\
    E_M(k) \\
  \end{array}
\right]=\mathcal{P}\left[
  \begin{array}{c}
    V(k) \\
    M(k) \\
  \end{array}
\right], \quad
\mathcal{P}=\left[\begin{array}{cc}
              B & B \\
              A+\lambda_1 & A+\lambda_2 \\
            \end{array}
          \right].
\label{transition matrix}
\end{eqnarray}
Then,
\begin{eqnarray}
\left[
  \begin{array}{c}
    V'(k) \\
    M'(k) \\
  \end{array}
\right]&=&P^{-1}\mathcal{M} P
\left[
  \begin{array}{c}
    V(k) \\
    M(k) \\
  \end{array}
\right]=\left[\begin{array}{cc}
              \lambda_1 & 0 \\
              0 & \lambda_2 \\
            \end{array}
          \right]
\left[
  \begin{array}{c}
    V(k) \\
    M(k) \\
  \end{array}
\right]\nonumber \\
\Rightarrow \left[
  \begin{array}{c}
    V(k) \\
    M(k) \\
  \end{array}
\right]&=&
\left[\begin{array}{c}
              V_0(k)e^{\lambda_1 t}\\ M_0(k)e^{\lambda_2 t}
         \end{array}
          \right].
\label{homogeneous_equations_diagonalized}
\end{eqnarray}
$\lambda_1$ and $\lambda_2$ are eigenvalues:
\begin{eqnarray}
\lambda_1&=&\frac{-(A+D)+\sqrt{(A+D)^2-4AD+4BC}}{2},\\
\lambda_2&=&\frac{-(A+D)-\sqrt{(A+D)^2-4AD+4BC}}{2}.
\label{eigenvalues}
\end{eqnarray}
We choose a leading term in each elements with the consideration of their coefficients $k\sim p$ or $k\sim r$. Then the coefficients are like
\begin{eqnarray}
A\sim 2\nu k^2,\quad B\sim b_1k^m,\quad C\sim c_2k^{m+2},
\quad D\sim d_2k^{v+2}.
\label{approximate_coefficients}
\end{eqnarray}
Here\footnote{The dimensional analysis of Eq.(\ref{Ev(t)}) implies $2m\sim v+2$ when the system is saturated.}, since $AD-BC\sim k^{v+4}-k^{2m+2}\sim 0$ as $k\rightarrow \infty$, the eigenvalues converge to $\lambda_1\sim 0,\quad \lambda_2\sim -2\nu k^2$ as `$k$' increases.\\

\noindent $E_V(k)$ and $E_M(k)$ are expressed like
\begin{eqnarray}
\left[
  \begin{array}{c}
    E_V(k) \\
    E_M(k) \\
  \end{array}
\right]=\left[\begin{array}{cc}
              b_1k^m & b_1k^m \\
              2\nu k^2& 0 \\
            \end{array}
          \right]\left[\begin{array}{c}
              V_0(k)e^{\sim 0\cdot t}\\ M_0(k)e^{-2\nu k^2 t}
         \end{array}
          \right].
\label{homogeneous_solutions1}
\end{eqnarray}
If `$V_0$' and `$M_0$' are replaced by $E_{V0}(k)\sim k^v$ and $E_{M0}(k)\sim k^m$, we find the saturated solutions:
\begin{eqnarray}
E_M(k)&=&2\nu k^2 V_0(k)\sim 2\nu k^2 \frac{E_{M0}(k)}{2\nu k^2}\sim k^m,\label{homogeneous_solutions EM2}\\
E_V(k)&=&b_1k^m\frac{E_{M0}(k)}{2\nu k^2}+
              b_1k^m\bigg[-\frac{E_{M0}(k)}{2\nu k^2}+\frac{E_{V0}(k)}{b_1k^m}\bigg]e^{-2\nu k^2 t}
              \sim k^{2m-2}
              \label{homogeneous_solutions EV2}.
\label{homogeneous_solutions 2}
\end{eqnarray}
For complete energy spectra, `$m$' is required. But it is difficult to pinpoint a representative magnetic power spectrum because $E_M(k)$ is a continuously changing curve. \cite{Schekochihin et al 2004} found `$m=0$', peak of $E_M(k)$, but we do not think a peak can be a representative power spectrum that drops continuously with the wavenumber `$k$'. On the other hand, \cite{Lazarian et al 2004} found `$m=-1$' using a simulation with a strong background magnetic field $\mathbf{b}_{ext}$, a filling factor, and balance relation $\textbf{B}\cdot \nabla \textbf{B} \sim \nu \nabla^2 \textbf{v}$. So we infer the index `$m=-1/2$' for the magnetic scaling factor in a system under the influence of a weak background magnetic field. Then, from Eq.(\ref{homogeneous_solutions EV2}) we get $E_V(k)$ `$k^{-3}$', which matches the simulation results well. Also this makes `$AD-BC\sim 0$', i.e., `$\lambda_1\sim 0$' and `$\lambda_2< 0$' so that the energy spectra become independent of time when they are saturated. If `$m=-1$' is chosen, `$E_V(k)\sim k^{-4}$' and $E_M(k)\sim k^{-1}$', coincident with the results of \citep{Cho et al 2003, Lazarian et al 2004}. A simple relation $E_M^2/E_V=k^2$ can be derived in high $Pr_M$.

\section{Discussion}
The analytic and simulation job in this paper are to realize the high $Pr_M$ plasma state in ICM. Magnetic field affects the fluid motion through Lorentz force $q\mathbf{v}\times \mathbf{B}$ to cause the rotational motion of ionized particles around the magnetic field. The effect of magnetic field, leading to the anisotropic system, competes with that of collision which transfers momentum to make the system isotropic. In fact the weakly collisional ICM plasma has few proper ways to constrict the anisotropic tendency. However, if background magnetic field is not too strong, a system driven by a random isotropic force in large scales eventually becomes isotropic overall although small scale eddies still tend to be anisotropic under the influence of magnetic field.\\

\noindent Simulation of high $Pr_M$ tells us some important features. The viscous scale $k_{\nu}$ is extended toward the much smaller diffusivity scale $k_{\eta}$, and nontrivial $E_V$ in this extended scale is a prerequisite to the growth of $E_M$. For the local and nonlocal energy transfer, besides energy gap in kinetic and magnetic eddies, additional specific geometrical relation between $\mathbf{v}$ and $\mathbf{B}$ is required. For example smaller $E_M(k)$ than $E_V(k)$ in larger scales boosts the nonlocal kinetic energy transfer from kinetic to magnetic eddies; and, smaller $E_V(k)$ than $E_M(k)$ in small scale helps the energy transfer from magnetic to kinetic eddies. However, the energy transfer in magnetic eddies is possible only when $\mathbf{B}\cdot\nabla \mathbf{V}$ (nonlocal transfer) or $-\mathbf{V}\cdot\nabla \mathbf{B}$ (local transfer) is nontrivial.
As a result of all these effects with the viscous damping, $E_V\sim k^{-3}$ and smoothly changing $E_M$ are finally saturated in the subviscous regime. Analytic analysis shows the viscous effect $\sim \nu k^2$ coupled with $E_M^2$ induces this unusual spectrum. (Eq.\ref{homogeneous_solutions EM2}, \ref{homogeneous_solutions EV2}).\\

\noindent Finally, as mentioned the anisotropic features of small scale cannot affect the large scale driven by the isotropic force. However, if there is an instability due to the anisotropic pressure in microscale, plasma distribution in the whole system may change. We have not discussed the influence of microscale instability due to the anisotropic pressure including viscosity and conductivity on the MHD system in this paper, but we will leave these important topics for the future research on ICM.

\section{Acknowledgements}
KP acknowledges support from the National Research Foundation of Korea through grant
2007-0093860. DP acknowledges the Korea Ministry of Education, Science and Technology, Gyeongsangbuk-Do and Pohang City for the support of the Junior Research Group at APCTP


\begin{appendix}

\section{Appendix}
\begin{figure}
\centering{
   \subfigure[]{
     \includegraphics[width=8cm]{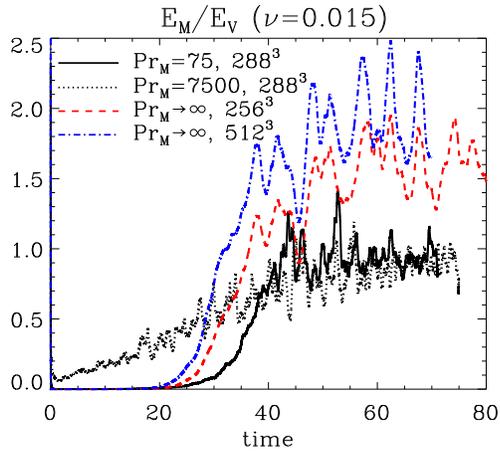}
     \label{f11}
    }
  }
\caption{The ratio $E_M$/$E_V$ of hyper diffusivity increases with resolution. In case of physical diffusivity the equipartition of $E_M$ and $E_V$ appears.}
\end{figure}

For `$A1$' in Eq.(\ref{momentum correlation equation in Fourier space}), we differentiate this triple correlation term over time to use Eq.(\ref{momentuem equation in Fourier space}), (\ref{magnetic induction equation in Fourier space}). Then, we have
\begin{eqnarray}
&&\big[\frac{\partial}{\partial t}+\nu\big(k^2+p^2+r^2\big) \big]\big\langle v_q(p)v_m(r)v_i(-k)\big\rangle_{A1}=\nonumber\\
&& \big\langle \big[\big(\frac{\partial}{\partial t}+\nu k^2\big)v_i(-k)\big]v_q(p)v_m(r)\big\rangle+\big\langle v_i(-k)\big[\big(\frac{\partial}{\partial t}+\nu p^2\big)v_q(p)\big]v_m(r)\big\rangle \nonumber\\
&&+\big\langle v_i(-k)v_q(p)\big[\big(\frac{\partial}{\partial t}+\nu r^2\big)v_m(r)\big]\big\rangle=\langle vvvv\rangle+\langle vvBB\rangle...
\label{tripple momentum correlation equation in Fourier space_A1}
\end{eqnarray}
If we see the first term, for example,
\begin{eqnarray}
\big\langle \big[\big(\frac{\partial}{\partial t}+\nu k^2\big)v_i(-k)\big]v_j(p)v_m(r)\big\rangle\delta_{p+r,k}=\sum_{j,l}
\big[M_{ins}(-k)\langle v_n(j)v_s(l)v_q(p)v_m(r)\rangle\nonumber\\
-M_{ins}(-k)\langle B_n(j)B_s(l)v_q(p)v_m(r)\rangle\big]\delta_{j+l,-k}.
\label{quadruple momentum correlation equation in Fourier space}
\end{eqnarray}
the differentiation generates the fourth order correlation. Another differentiation just induces the fifth order correlation. So we need an assumption to close this equation. It is known that statistically turbulent quantities follow a normal distribution. And the fourth-order term $\langle uuuu \rangle$ can be decomposed into the combination of second-order correlation terms like below: quasi-normal approximation \citep{Proudman et al 1954, Tatsumi 1957}:
\begin{eqnarray}
\langle u_1u_2u_3u_4\rangle=\langle u_1u_2\rangle\langle u_3u_4\rangle+\langle u_1u_3\rangle\langle u_2u_4\rangle+\langle u_1u_4\rangle\langle u_2u_3\rangle.
\label{Quasi normal fourth order}
\end{eqnarray}
So if these second order correlation terms are replaced by energy spectrum expressions as follows:
\begin{eqnarray}
4\pi k^2\langle v_i(k)v_q(k')\rangle&=&P_{iq}(k)E_V(k)\delta_{k+k',0},\\
4\pi k^2\langle B_i(k)B_q(k')\rangle&=&P_{iq}(k)E_M(k)\delta_{k+k',0},\nonumber\\
4\pi k^2\langle B_i(k)v_q(k')\rangle&=&P_{iq}(k)H_{BV}(k)\delta_{k+k',0}.\nonumber\\
(P_{iq}=\delta_{iq}-\frac{k_ik_q}{k^2})\nonumber
\label{Second order velocity magnetic field correlation}
\end{eqnarray}
Eq.(\ref{tripple momentum correlation equation in Fourier space_A1}) can be rewritten like
\begin{eqnarray}
&&\big[\frac{\partial}{\partial t}+\nu\big(k^2+p^2+r^2\big) \big]\big\langle v_q(p)v_m(r)v_i(-k)\big\rangle_{A1}=\nonumber\\
&&2M_{ins}(-k)\sum_{p,r} (4\pi p^2)^{-1}(4\pi r^2)^{-1}P_{nq}(p)P_{ms}(r)[E_V(p)E_V(r)-H_{BV}(p)H_{BV}(r)]+\nonumber\\
&&2M_{qns}(p)\sum_{p,r} (4\pi k^2)^{-1}(4\pi r^2)^{-1}P_{in}(k)P_{ms}(r)[E_V(k)E_V(r)-H_{BV}(k)H_{BV}(r)]+\nonumber\\
&&2M_{mns}(r)\sum_{p,r} (4\pi k^2)^{-1}(4\pi r^2)^{-1}P_{in}(k)P_{js}(p)[E_V(k)E_V(p)-H_{BV}(k)H_{BV}(p)]\nonumber\\
&&\equiv L^{vv1}_{iqm}(k,p,r;t).
\label{tripple momentum correlation equation in Fourier space_A1_result}
\end{eqnarray}
However, when the fourth-order correlation is decomposed into the combinations of second order terms, the summation of decomposed ones, i.e., right hand side of Eq.(\ref{Quasi normal fourth order}), can be larger than its actual value. This can cause a negative energy spectrum, which cannot be allowed \citep{Ogura 1963}. So \citep{Orszag 1970} introduced an eddy damping coefficient $\mu_{kpr}$ of which dimension is `$\sim 1/t$'. Its more detailed expression \citep{Pouquet et al 1976} can be contrived, but the dimension `$\sim 1/t$' does not change. We assume it to be sort of a reciprocal of time constant for simplicity in this paper. Then, with a simple integration we can find the third correlation term:
\begin{eqnarray}
\big\langle v_q(p)v_m(r)v_i(-k)\big\rangle_{A1}=\int^t e^{-(\nu(k^2+p^2+r^2)+\mu_{kpr})(t-\tau)}L^{vv1}_{iqm}(k,p,r;\tau)d\tau.
\label{tripple momentum correlation equation calculation_V2}
\end{eqnarray}
We can also calculate the representation of `$A3$' in the same way:
\begin{eqnarray}
\big\langle B_q(p)B_m(r)v_i(-k)\big\rangle_{A3}=\int^t e^{-(\nu k^2+\mu_{kpr})(t-\tau)}L^{vv3}_{iqm}(k,p,r;\tau)d\tau.
\label{tripple momentum correlation equation calculation_V3}
\end{eqnarray}
Since `$A2$' and `$A4$' are `-$A1$' and `-$A3$' respectively, Eq.(\ref{momentum correlation equation in Fourier space}) are
\begin{eqnarray}
\big(\frac{\partial}{\partial t}+2\nu k^2 \big)\langle v_i(k)v_i(-k)\rangle=\sum_{p+r=k}2M_{iqm}(k)\bigg[\int^t e^{-(\nu(k^2+p^2+r^2)+\mu_{kpr})(t-\tau)}L^{vv1}_{iqm}(k,p,r;\tau)d\tau\nonumber\\
-\int^t e^{-(\nu k^2+\mu_{kpr})(t-\tau)}L^{vv3}_{iqm}(k,p,r;\tau)d\tau\bigg].\nonumber\\
\label{kinetic energy equation in Fourier space final form}
\end{eqnarray}
If $L^{vv1}_{iqm}(k,p,r;\tau)$ or $L^{vv3}_{iqm}(k,p,r;\tau)$ is larger than $(\nu(k^2+p^2+r^2)+\mu_{kpr})^{-1}$ or $(\nu k^2+\mu_{kpr})^{-1}$, this equation can be markovianized. Thus, with the definition of a triad relaxation time $\Theta(t)$ \citep{Frisch et al 1975} we get
\begin{eqnarray}
&&\big(\frac{\partial}{\partial t}+2\nu k^2 \big)E_V(k)=\nonumber\\
&&\sum_{p+r=k}4\pi k^2M_{iqm}(k)\bigg[\bigg(\frac{1-e^{-(\nu(k^2+p^2+r^2)+\mu_{kpr})t}}{\nu(k^2+p^2+r^2)+\mu_{kpr}}\bigg)
L^{vv1}_{iqm}(k,p,r;t)\nonumber\\
&&-\bigg(\frac{1-e^{-(\nu k^2+\mu_{kpr})t}}{\nu k^2+\mu_{kpr}}\bigg)
L^{vv3}_{iqm}(k,p,r;t)\bigg]\nonumber\\
&&\equiv\sum_{p+r=k}4\pi k^2M_{iqm}(k)\bigg[\Theta^{\nu\nu\nu}_{kpr}(t)
L^{vv1}_{iqm}(k,p,r;t)-\Theta^{\nu\eta\eta}_{kpr}(t)L^{vv3}_{iqm}(k,p,r;t)\bigg].\nonumber\\
\label{Markovianized kinetic energy equation in Fourier space final form}
\end{eqnarray}
Using a trigonometric relation: $\mathbf{k}\cdot \mathbf{p}=kpz$, $\mathbf{k}\cdot \mathbf{r}=kry$, $\mathbf{p}\cdot \mathbf{r}=-prx$ (Fig.\ref{f11}) and $d\mathbf{p}\,d\mathbf{r}$=$\frac{2\pi p\,r}{k}dp\,dr$, we can simplify this expression:
\begin{eqnarray}
\frac{1}{2}\int dp\,dr\,\Theta^{\nu\nu\nu}_{kpr}(t)\frac{k^3}{p\,r}
(1-2y^2z^2-xyz)[E_V(p)E_V(r)-H_{BV}(p)H_{BV}(r)].
\label{coefficient_calculation_Ev first}
\end{eqnarray}
Eq.(\ref{EDQNMEv(t)}), (\ref{EDQNMEm(t)}) can be derived using a similar way.

\end{appendix}

\begin{thebibliography}{}

\bibitem[Batchelor (1950)]{Batchelor 1950}
Batchelor, G. K. 1950, Proc. Roy. Soc. London, Ser. A, 201, 405

\bibitem[Bovino et al (2013)]{Bovino et al 2013}
Bovino, S., Schleicher, D., R., G., \& Schober, J., 2013, New J. Phys., 15, 013055

\bibitem[Brandenburg (2001)]{Brandenburg 2001}
Brandenburg, A. 2001, \apj, 550, 824\\

\bibitem[Chew et al.(1956)]{Chew et al 1956}
Chew, G., F., Goldberger, M., L., \& Low, F., E., 1956, RSPSA, 236, 112

\bibitem[Cho et al.(2003)]{Cho et al 2003}
Cho, J., Lazarian, A., \& Vishniac, E., 2003, \apj, 595, 812

\bibitem[Cho et al.(2009)]{Cho et al 2009}
Cho, J., Vishniac, E., Beresnyak, A., Lazarian, A., \& Ryu, D., 2009, \apj,  693, 1449

\bibitem[Cho \& Yoo(2012)]{Cho and Yoo 2012}
Cho, J., \& Yoo, H., 2014, \apj, 780, 99

\bibitem[Cho(2014)]{Cho 2014}
Cho, J., 2014, 797, 133

\bibitem[Frisch et al.(1975)]{Frisch et al 1975}
Frisch, U. , Pouquet, A., Leorat, J. \& Mazure, A., 1975, J. Fluid Mech., 68, 769

\bibitem[Jones(2008)]{Jones 2008}
Jones, T. W., 2008, Astronomical Society of the Pacific Conference Series, 386, 398


\bibitem[Kazantsev(1968)]{Kazantsev 1968}
Kazantsev, A., P., 1968, JETP, 26, 1031

\bibitem[Kleeorin et al.(1996)]{Kleeorin et al 1996}
Kleeorin, N., Mond, M., \& Rogachevskii, I., 1996, \aap, 307, 293

\bibitem[Kolmogorov(1941)]{Kolmogorov 1941}
Kolmogorov, A., 1941, Akademiia Nauk SSSR Doklady, 30, 301

\bibitem[Kraichnan \& Nagarajan(1967)]{Kraichnan and Nagarajan 1967}
Kraichnan R. H. \& Nagarajan, S., 1967, Physics of Fluids, 10, 859

\bibitem[Krause \& R\"adler(1980)]{Krause and Radler 1980}
Krause, F. \& R\"adler, K. H., 1980, Mean-field magnetohydrodynamics and dynamo theory

\bibitem[Kulsrud and Anderson (1992)]{Kulsrud and Anderson 1992}
Kulsrud, R. M. \& Anderson, S. W. 1992, ApJ, 396, 606

\bibitem[Lazarian et al.(2004)]{Lazarian et al 2004}
Lazarian, A., Vishniac, E. T., \& Cho, J., \apj, 2004, 603, 180


\bibitem[Leslie \& Leith(1975)]{Leslie and Leith 1975}
Leslie, D. C. \& Leith, C. E., 1975, Physics Today, 28, 59


\bibitem[McComb(1990)]{McComb 1990}
McComb, W. D., 1990, The physics of fluid turbulence

\bibitem[Moffatt(1978)]{Moffatt 1978}
Moffatt, H. K., 1978, Magnetic Field Generation in Electrically Conducting Fluids

\bibitem[Narayan \& Medvedev(2001)]{Narayan and Medvedev 2001}
Narayan, R., \& Medvedev, M. V., 2001, \apjl, 562, L129

\bibitem[Ogura(1963)]{Ogura 1963}
Ogura, Y., 1963, J. Fluid Mech., 16, 33

\bibitem[Orszag(1970)]{Orszag 1970}
Orszag, S. A., 1970, J. Fluid Mech., 41, 363

\bibitem[Park(2013)]{Park 2013}
Park, K., 2013, \mnras, 434, 2020

\bibitem[Park \& Blackman(2012a)]{Park and Blackman 2012a}
Park, K., \& Blackman, E. G., 2012a, \mnras, 419, 913

\bibitem[Park \& Blackman(2012b)]{Park and Blackman 2012b}
Park, K., \& Blackman, E. G., 2012b, \mnras, 423, 2120

\bibitem[Pouquet et al.(1976)]{Pouquet et al 1976}
Pouquet, A., Frisch, U., \& Leorat, J., 1976, J. Fluid Mech., 77, 321

\bibitem[Proudman et al.(1954)]{Proudman et al 1954}
Proudman, I., \& Reid, W. H., 1954, Royal Society of London Philosophical Transactions Series A, 247, 163

\bibitem[Ruzmaikin et al.(1982)]{Ruzmaikin et al 1982}
Ruzmaikin, A. A., \& Shukurov, A. M., 1982, Astrophysics and Space Science, 82, 397

\bibitem[Ryu et al.(2008)]{Ryu et al 2008}
Ryu, D., Kang, H., Cho, J., \& Das, S., 2008, SCIENCE, 320, 909

\bibitem[Ryu et al.(2012)]{Ryu et al 2012}
Ryu, D., Porter, D. H., Cho, J., \& Jones, T., W., 2012, AAS...21933803R

\bibitem[Santos-Lima et al.(2011)]{Santos-Liman et al 2011}
Santos-Lima, R., de Gouveia Dal Pino, E., M., Falceta-Gon{\c c}alves, D., Lazarian, A., \& Kowal, G., 2011, IAUS, 274, 482

\bibitem[Santos-Lima et al.(2014)]{Santos-Liman et al 2014}
Santos-Lima, R., de Gouveia Dal Pino, E., M., Kowal, G., Falceta-Gon{\c c}alves, D., Lazarian, A., \& Nakwacki, M., S., 2014, \apj, 781, 84

\bibitem[Schekochihin et al.(2002)]{Schekochihin et al 2002}
Schekochihin, A. A., Maron, J. L., Cowley, S. C., \& McWilliams, J. C., 2002, \apj, 576, 806

\bibitem[Schekochihin et al.(2004)]{Schekochihin et al 2004}
Schekochihin, A. A., Cowley, S. C., Taylor, S. F., Maron, J. L., \& McWilliams, J. C., 2004, \apj, 612, 276

\bibitem[Schekochihin et al.(2005)]{Schekochihin et al 2005}
Schekochihin, A. A., Cowley, S. C., Kulsrud, R. M., Hammett, G. W., \& Sharma, P., 2005, \apj, 629, 139

\bibitem[Schekochihin et al.(2005)]{Schekochihin et al 2005}
Schekochihin, A. A., Cowley, S. C., Kulsrud, R. M., Hammett, G. W., \& Sharma, P., 2005, \apj, 629, 139


\bibitem[Schekochihin et al.(2008)]{Schekochihin et al 2008}
Schekochihin, A. A., Cowley, S. C., Kulsrud, R. M., Rosin, M. S., \&  Heinemann, T., 2008, Physical Review Letters, 100(8), 081301

\bibitem[Schekochihin et al.(2010)]{Schekochihin et al 2010}
Schekochihin, A. A., Cowley, S. C., Rincon, F., \$ Rosin, M. S., 2010, \mnras, 405, 291

\bibitem[Schober et al.(2012)]{Schober et al 2012}
Schober, J., Schleicher, D., Federrath, C., Klessen, R., \& Banerjee, R., 2012, Phys. Rev. E., 85, 026303

\bibitem[Tatsumi(1957)]{Tatsumi 1957}
Tatsumi, T., 1957, Royal Society of London Proceedings Series A, 239, 16

\bibitem[Vogt \& En{\ss}lin (2003)]{Vogt and Ensslin 2003}
Vogt, C., \& En{\ss}lin, T., A., 2003, A\&A, 412, 373

\bibitem[Yoshizawa(2011)]{Yoshizawa 2011}
Yoshizawa, A., 2011, Hydrodynamic and Magnetohydrodynamic Turbulent Flows: Modelling and Statistical Theory (Fluid Mechanics and Its Applications

\bibitem[Yousef et al.(2007)]{Yousef et al 2007}
Yousef, T., A., Rincon, F., \& Schekochihin, A., A., 2007, JFM, 575, 111


\end{thebibliography}
\end{document}